\documentclass[aps, pre, twocolumn, longbibliography,superscriptaddress]{revtex4-2}
\usepackage{bm,amsmath,graphicx, enumerate, mathtools,multirow}
\usepackage[colorlinks=true,linkcolor=MidnightBlue,urlcolor=black,citecolor=MidnightBlue,anchorcolor=MidnightBlue]{hyperref}
\usepackage[dvipsnames]{xcolor}
\begin{document}
%%%%%====================================
%%%%%====================================
%\title{Roto-translational interfacial fluctuations in a self-propelling colloidal interface}
% \title{Topology-specific growth and roughening in a self-interacting propelled colloidal interface}
% \title{Growth and roughening of an active interface: role of topology and self-interactions}
% \title{Dynamics and fluctuations of an active interface: role of topology and self-interactions}
% \title{Fluctuations of an active interface: role of topology and phoretic interactions}
% \title{Fluctuations of an active colloidal interface: role of self-interactions and roto-translational coupling}
\title{Shape-specific fluctuations of an active colloidal interface}
\author{Arvin Gopal Subramaniam}
\thanks{ph22d800@smail.iitm.ac.in}
\affiliation{Department of Physics, Indian Institute of Technology Madras, Chennai, India}
\affiliation{Center for Soft and Biological Matter, IIT Madras, Chennai,  India}
\author{Tirthankar Banerjee}
\thanks{tirthankar.banerjee@uni.lu}
\affiliation{Department of Physics and Materials Science, University of Luxembourg, Luxembourg, Luxembourg }
\author{Rajesh Singh}
\thanks{rsingh@physics.iitm.ac.in}
\affiliation{Department of Physics, Indian Institute of Technology Madras, Chennai, India}
\affiliation{Center for Soft and Biological Matter, IIT Madras, Chennai,  India}

\begin{abstract}
%Inspired by a recently synthesized class of active interfaces consisting of linked self-propelled colloids, we study the dynamics and scaling exponents of a self-chemically interacting active interface. We characterize the different possible steady-state shapes of the interface in the space of model parameters and show that there exists a parameter regime where the steady-state profile propels deterministically with a finite curvature. We show that the height fluctuations in this phase fall into the Family-Vicsek scaling class, with novel dynamical and roughness exponents. A novel negative roughness exponent (``smoothness": a decrease in steady-state fluctuations with system size) for the orientation fluctuations are also reported.  

Motivated by a recently synthesizable class of active interfaces formed by linked self-propelled colloids, we investigate the dynamics and fluctuations of a phoretically (chemically) interacting active interface with roto-translational coupling. We enumerate all steady-state shapes of the interface across parameter space and identify a regime where the interface acquires a finite curvature, leading to a characteristic “C-shaped” topology, along with persistent self-propulsion. In this phase, the interface height fluctuations obey Family–Vicsek scaling but with novel exponents: a dynamic exponent $z_h \approx 0.5$, a roughness exponent $\alpha_h \approx 0.9$ and a super-ballistic growth exponent $\beta_h \approx 1.7$. In contrast, the orientational fluctuations of the colloidal monomers exhibit a negative roughness exponent, reflecting a surprising {\em smoothness law}, where steady-state fluctuations diminish with increasing system size. 
Together, these findings point towards a unique non-equilibrium universality class associated with self-propelled interfaces of non-standard shape.
%Furthermore, distinct scaling exponents are uncovered in the locally flat regime, where monomer-level averages are taken within flat subsets of the interface. 
%Addition of inter-monomeric non-equilibrium chemical interactions leads to a deviation from this distribution, with the fluctuation exponents unchanged. Steady-state chain length scaling exponents are also reported.
\end{abstract}
\maketitle

\maketitle

\section{Introduction}\label{sec:intro}
Out-of-equilibrium agents with suitably chosen interactions are known to exhibit emergent collective order, most prominently in the form of a global polar order ~\cite{toner2024physics, vicsek1995novel}.
Such order can arise from alignment rules \cite{vicsek1995novel, chate2008collective}, topological interactions \cite{ballerini2008interaction}, long-ranged attraction or repulsion \cite{caprini2023flocking, subramaniam2025minimal}, and/or other generic behavioral couplings \cite{grossmann2013self}. 
%On the other hand, in a separate field of statistical physics, the physics of growing interfaces is well established, with various well-known results on statistical fluctuations of the interface height  \cite{kardar1998nonequilibrium, barabasi1995fractal, krapivsky2010kinetic, dean2025exact}. 
In parallel, a distinct body of work in statistical physics has established a deep understanding of fluctuating interfaces, where the statistical properties of the interface height obey universal scaling laws \cite{kardar1998nonequilibrium, barabasi1995fractal, krapivsky2010kinetic, dean2025exact}.
A cornerstone result in this field is the Family–Vicsek (FV) scaling law, which relates the roughness of the interface to the system size and time through universal exponents \cite{vicsek1984dynamic, fujimoto2020family}.
%The canonical scaling law uncovered here is  Family-Vicsek (FV) scaling law characterized by dynamic and roughness exponents \cite{vicsek1984dynamic, fujimoto2020family}, 
%with established theoretical results based on two solvable models, namely the Edwards-Wilkinson (EW) and Kardar-Parisi-Zhang (KPZ) models 
These exponents are known for certain solvable models, such as the Edwards–Wilkinson (EW) and Kardar–Parisi–Zhang (KPZ) equations
\cite{kardar1998nonequilibrium, barabasi1995fractal, corwin2012kardar},
while numerical and experimental studies have revealed a wealth of deviations, suggesting novel universality classes~\cite{family1985scaling, kim1991surface, sarma1991new, degawa2006distinctive}. Bridging these two domains, the study of {\em active interfaces}, i.e., interfaces driven out of equilibrium by internal activity, is a subject of recent interest ~\cite{cagnetta2018active, goutaland2021binding,adkins2022dynamics, caballero2025, cates2025active, maire2025conservation}, raising new questions about how activity reshapes interfacial fluctuations and scaling behavior.
%with various deviations from these (including hypothetical new universality classes) reported in simulations (e.g. \cite{family1985scaling, kim1991surface, sarma1991new}) and experiment (e.g. \cite{degawa2006distinctive}). Adding a non-equilibrium nature to interfacial problems - so-called active interfaces - is an area of recent interest \cite{cagnetta2018active, goutaland2021binding,adkins2022dynamics, caballero2025, cates2025active}. \\

In this article, we study a polar colloidal chain as a non-equilibrium interface in $1+1$ dimensions. This is inspired by recent experiments that have been able to synthesize an autophoretic polar chain via chemical self-interactions \cite{kumar2023emergent}, as opposed to other propulsion mechanisms via external actuation \cite{nishiguchi2018flagellar, snezhko2011magnetic, biswas2017linking}. We study the dynamical steady-states of this system and show that within a specific parameter regime the chain ballistically propels with a deterministic ``C-shape"\cite{kumar2023emergent, subramaniam2024rigid}. As opposed to conventional growing (circular) interfaces, the non-equilibrium interface we study here displays the additional phenomena (in addition to being propelled) of curvature acquisition during its dynamics, attained via the inter-monomeric phoretic interactions, thus adding additional time scales to the conventional early time deterministic plus late time diffusive behaviour \cite{wu2000particle}. The polar nature of the interface of this topology (and hence the reason it breaks translational symmetry and propels in a given direction) enables calculations of height fluctuations of the monomers about a mean height defined across the interface.

We report that the interface-height fluctuations of this chain obey FV scaling, characterized by a set of dynamic ($z_h \approx 0.5$) and roughness ($\alpha_h \approx 0.9$) exponents. Correspondingly, the fluctuations preceding the steady state are captured by a super-ballistic growth exponent $\beta_h \approx 1.7$. The steady-state orientational fluctuations of the colloidal monomers in the C-shape phase, in contrast, decrease with the chain length, exhibiting a \textit{negative} roughness (i.e. ``smoothness") exponent ($\alpha_\theta \approx -0.5$). Moreover, we report a distinct set of scaling exponents characterizing the locally flat regime, defined by monomer-level averages restricted to flat portions of the chain and, independently, by taking the infinite-chain limit.
%Moreover, a separate set of scaling exponents are reported for the \textit{locally flat} regime, where monomer averages are taken within a flat subset of the chain and (separately) as the chain length goes to infinity. 
Given that the shape of the propelling interface does not fall under conventionally studied circular or flat interfaces \cite{takeuchi2018appetizer}, these exponents thus constitute a unique non-equilibrium signature of an interface with a stereotypic ``C-shape" topology.

\begin{figure*}[t!]
    \includegraphics[width=0.88\textwidth]{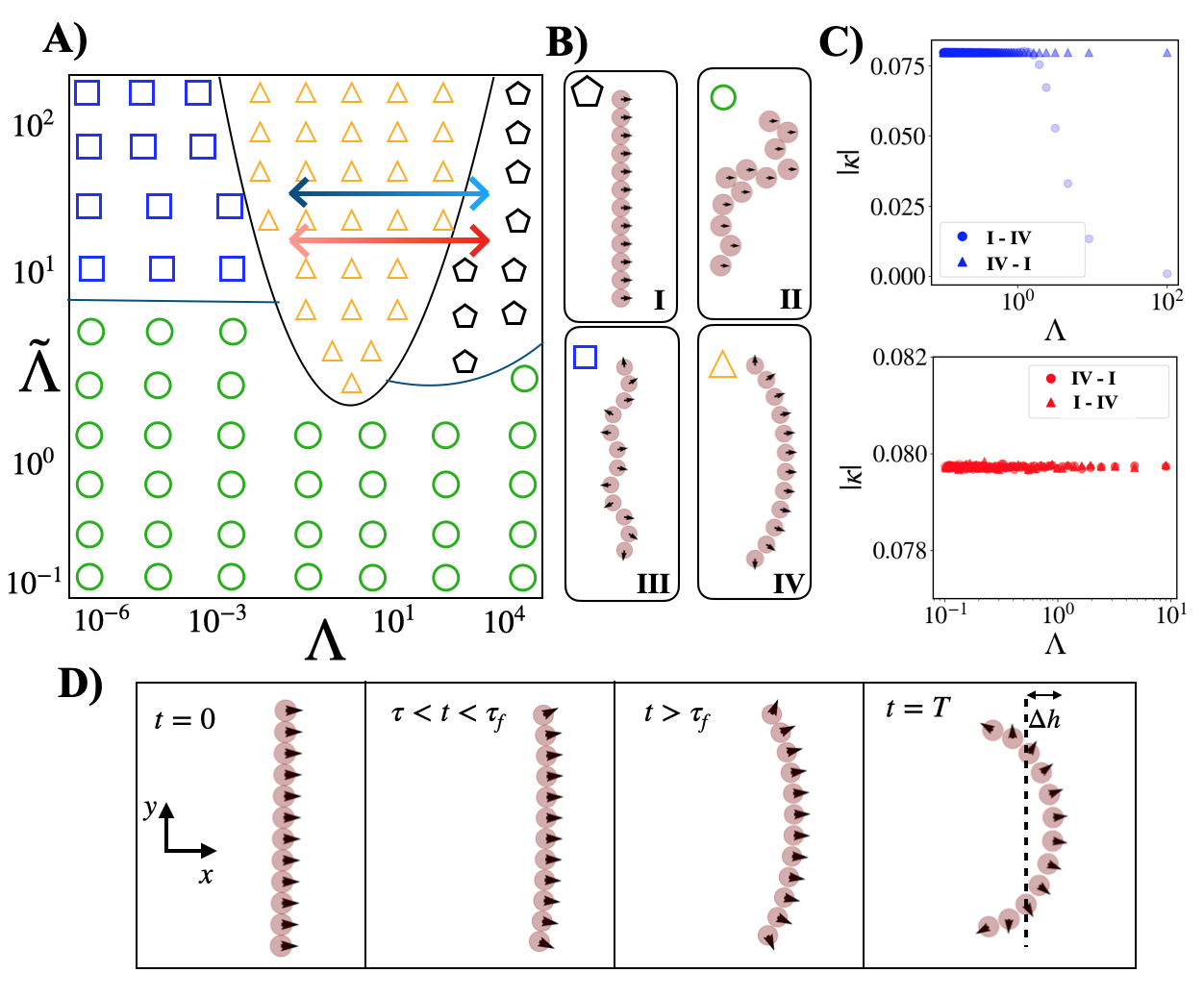}
    \caption{
    \textbf{A}) Phase diagram in $\Lambda$-$\tilde\Lambda$ plane, see Eq.\eqref{eq:peclet_nums}. 
    The log absolute value of curvature has been used to delineate the phases. The phase diagram has been drawn with for a chain with $N=256$ number of monomers. \textbf{B}) Representative images for each phase is shown along with the marker key for a smaller chain for clarity. Values of $(\Lambda, \tilde{\Lambda})$ 
    are \textbf{I}: $(10^{4},10^{2})$, 
    \textbf{II}: $(10^{1},10^{0})$, 
    \textbf{III}:  $(10^{-3},10^{2})$, 
    \textbf{III}: $(10^{1},10^{2})$. 
    \textbf{C}) Displays the effect of selection statistics of the C-shape under two separate protocols, where the absolute value of the curvature $| \kappa |$ has been plotted against the control parameter. 
    \textit{Top} corresponds to that of variation of 
    $\Lambda$ starting in Phase I, whilst \textit{bottom} is starting in Phase IV. Note that $| \kappa| \approx 0.0797$ indicates the C-shape absolute curvature. Light points denote the beginning of the protocol whilst dark the end for the forward direction, with the opposite color shading for the reverse. Color shade and parameter values correspond to arrows shown in panel \textbf{A} for transitions between states \textbf{I} and \textbf{IV} of the phase diagram. \textbf{D}) An example snapshot of the evolution in Phase IV with $N=12$ and $\tau_f = 0.2$. The rightmost snapshot shows the steady-state shape with an example mean height (vertical dotted line) and definition of $\Delta h$ in Eq.(\ref{eq:d2h}). }
    \label{fig_phase}
\end{figure*}

 The paper is structured as follows. In Sect.~\ref{sec:2}, we introduce the model and enumerate the relevant length and time scales of our model. In Sec.~\ref{sec:3} , we study the dynamical regimes of the model as functions of selected dimensionless numbers and obtain a phase diagram [Fig.~\ref{fig_phase}]. The main results of this paper are presented in Sec.~\ref{sec:chemo-rep}, where we discuss in detail the height and orientational fluctuations of this system, in addition to the effect of finite rotational noise. We finally discuss the significance of our findings and future directions in Sec.~\ref{sec:summ}.

%% --------------------------

\section{Model}\label{sec:2}
%\subsection{Particle-based model for chemically-interacting chain}
We consider an active polymer consisting of chemically interacting active particles as monomers.  The $i$th active particle - centered at $\mathbf r_i=(x_i,y_i)$ - is confined to move in two-dimensions. It self-propels with a speed $v_s$,  along the directions $\mathbf e_i =(\cos\theta_i,\,\sin\theta_i)$.
 The position and orientation of the $i$th particle are determined by the following evolution equations:
 \begin{subequations}
\begin{align}
\dot{\mathbf {r}}_i &  = v_s\, \mathbf e_i  + \mu \,\mathbf F_i^{b}
   + \sqrt{2D_t}\,\bm\xi^t_{i},\\
\dot{\theta}_i  & = \chi_r\left(\mathbf e_i\times \mathbf { J}_i
    \right)
   + \sqrt{2D_r}\,\bm\xi^r_{i}. 
\end{align}
\label{eq:dyn}     
 \end{subequations}
In the above equations, $\mu$ is mobility, $D_t$ and $D_r$, respectively, are translational and rotational diffusion constants of the particle, while $\bm \xi^t$ and $\bm \xi^r$  are white noises with zero mean and unit variances. The constant $\chi_r$ is taken to be positive so that the particles rotate away from each other (chemo-repulsive). The force on the $i$th particle is given as:  
$\mathbf F_i^{b} = -{\partial U}/{\partial \mathbf r_i}$, 
while we have:
$  
U= \sum^{N-1}_{i=1}\mathcal U^b   (\mathbf r_i,\mathbf r_{i+1})$.  
Here, $\mathcal U^b(\mathbf r_i,\mathbf r_{j})=k_b\left(r_{ij}-r_0\right)^2$ is a harmonic potential of stiffness $k_b$ and natural length $r_0$ which holds the chain together, while $r_{ij}=|\mathbf {r}_i - \mathbf {r}_j|$. 

The orientation of the particles, given by the angle $\theta_i$, changes due to coupling to a chemical (phoretic) field $c({\bf r},t)$. The chemical interactions between the monomers of the chain are contained in $\mathbf { J}_i =- \partial c/\partial \mathbf {r}_i$, where $c$ is the concentration of the phoretic field. It is worthwhile to note that the positional and orientational dynamics are coupled in our model. This roto-translational coupling leads to rich phenomenology of our model, as we describe below.  
In the steady-state, the solution of the concentration profile $c(\bm r)$ follows from the equation: $D_{c}\nabla^2 c(\boldsymbol r, t) + \sum_{i=1}^Nc_0\, \delta(\boldsymbol r- \boldsymbol r_i) = 0$, where $D_{c}$ is the diffusion coefficient of the filled micelles and $c_0$ is emission constant of the micelles. This gives the expression for the current:
\begin{align}
    \mathbf{J}_{i} = \frac{c_0}{4\pi D_c}\sum_{\substack{j=1\\ j\neq i}}^N \frac{\mathbf{r}_{i} - \mathbf{r}_{j}}{|\mathbf{r}_{i} - \mathbf{r}_{j}|^{3}}
\label{eq:chem_field}
\end{align}
This current is thus interpreted as that which is instantaneously deposited on the centers of the colloids. 

Variants of the above model have been studied recently \cite{kumar2023emergent,subramaniam2024rigid}; a detailed phase diagram of the model was obtained in \cite{subramaniam2024rigid} in terms of the dimensionless group of the system. Here, we focus on the fluctuations in the so-called ``C-shape" phase, where the chain spontaneously acquires a stereotypic shape resembling the alphabet C, and propels in a direction normal to its tangent \cite{kumar2023emergent, subramaniam2024rigid}. We note that the ``C-shape" topology we refer to is specific to systems in which roto-translational coupling in the dynamics induces the steady-state shape. The system we study being ``dry" with such a coupling, is distinct from other similar topologies seen in (for instance) the physics of sedimenting filaments \cite{li2013sedimentation}. In the limit of $N\rightarrow \infty$ it corresponds to an interface which is moving. A schematic of this is shown in Fig.~\ref{fig_phase}D. We focus on the fluctuations of that moving interface in this paper. 

For the entire paper (unless specified otherwise), we use the following initial conditions:
\begin{subequations}
\label{eq:init_cond}
\begin{align}
    \theta_{i}(0) &= 0, \qquad x_{i} (0)= 0, \qquad \forall i,
    \qquad\qquad \\ 
    y_{i+1}(0) &= y_{i}(0) + 2b, \qquad\qquad\,\,\,
    1 \leq i \leq N-1
\end{align}
\end{subequations}
where $b$ is the radius of each colloid. Note that $r_0 = 2b$. 
All numerical results in this paper are generated with an explicit Euler-Maruyama integrator with time step $dt=0.01$ and simulation time $T=7.2 \times 10^{5}$. For all simulations here (unless stated otherwise) $b=1$, $r_0=2b$, $c_0=1$, $D_c = 1$, $\hat{\mu} = \mu k_{b} = 91.6$, and $v_s = 1$. $D_r=0$ for all sections, with the exception of Sec. \ref{sec:VC} where values are explicitly stated. Wherever realization average is performed, we use $50$ realizations, with the exception of $10$ used in Appendix \ref{sec:hist_dep}. 
%\textcolor{black}{Why is $\hat \mu$ mentioned with a unit while others are not?}
%\\
%We first compute the fluctuations using the above particle-based model. We then present a minimal continuum model and 

%We show that in the early time regime, i.e., before the formation of any curvature in the bulk,  the fluctuations of the interface height are reminiscent of those of an Edwards-Wilkinson interface driven by a constant velocity. An additional term is needed to explain the final steady-state when the chain acquires a curvature due to activity. Thus, we show that that activity induced via roto-translational coupling brings a new scaling exponent for a growing interface. \\

%\subsubsection{Dimensionless numbers}

\section{Dynamical regimes}\label{sec:3}

We first define some typical time scales, ratios of which determine the relevant dimensionless numbers needed to understand the non-equilibrium phase diagram (Fig.~\ref{fig_phase}):
\begin{align}
    \tau = \frac{b}{v_{s}}, \quad
    \tau_{t} = \frac{b^{2}}{D_{t}}, \quad
    \tau_f = \frac{b^{3}}{\chi_r}, \quad
    \tau_r = \frac{1}{D_r}.
\label{eq:tscales}
\end{align} 
 Here, $\tau$ is a spontaneous propulsion time scale, which is the time it takes for an isolated particle to move a distance equaling its radius in absence of any reorientation, $\tau_f$ sets the average time during which the orientation of the colloid changes in response to the chemical field. $\tau_t$ and $\tau_f$ are the time scales set by the translation noise $D_t$ and the rotational noise $D_r$, respectively. Competition between these time scales gives rise to different dynamical and scaling regimes of the propelling interface.

We define two dimensionless activity parameters:
\begin{align}
\Lambda =\frac{\tau_{f}}{\tau}= \frac{b^2\,v_{s}}{\chi_r}
,\qquad
    \tilde\Lambda &= \frac{\tau_{t}}{\tau}= \frac{b v_s}{ D_{t}}.
%     \\
%     \mathrm{Pe}^{3} &= \frac{\tau_{r}}{\tau}= \frac{v_s }{bD_r},
% \qquad
%     \mathrm{Pe}^{4} =\frac{\tau_{t}}{\tau_f}= \frac{\chi_r}{bD_t},
\label{eq:peclet_nums}
\end{align}
%In addition, we can define dimensionless number such as  $\mathcal{A} =\frac{\tau_{f}}{\tau}= \frac{b^2\,v_{s}}{\chi_r}$
The dimensionless number $\Lambda$ quantifies the competition between deterministic rotations (due to phoretic interactions) and deterministic propulsion. The dimensionless number $\tilde \Lambda$ quantifies deterministic and random motion in the positional sector. A variant on the above model, proposed in section \ref{sec:2}, was introduced in \cite{kumar2023emergent} and studied further in \cite{subramaniam2024rigid}.
In both these papers, we had considered the role of trail created by the particles. Here, we first assume instantaneous chemical interactions, and discuss the effect of trail-mediated history later (in Appendix~\ref{sec:hist_dep}). 
Additionally, we note that the previous works ignored the role of translational noise $D_t$. In \cite{subramaniam2024rigid}, it was found that a competition between $\tau_{r}$ and the chemical diffusivity of trails leads to the formation of a stable C-shape chain. \\

It is informative to study the model described above in terms of the two dimensionless numbers in Eq.~\eqref{eq:peclet_nums}. In Fig. \ref{fig_phase}, we present a phase diagram in the plane of $\Lambda$  and $\tilde\Lambda$. We find that the C-shape is sustained for a selected range of dimensionless numbers $\Lambda$ and $\tilde\Lambda$ \cite{kumar2023emergent, subramaniam2024rigid}. Let us construct in addition a fluctuating length scale $l_f = \frac{D_t}{v_s}$ and a correlation length scale $l_C = \frac{\chi_r}{b v_s}$ to aid with the analysis. First, in the limit of $\Lambda \gtrsim 10^{2}, \tilde{\Lambda} \gtrsim 10^1$, there exists insufficient rotations and the chain remains flat within the simulation time scale \cite{subramaniam2024rigid}. This corresponds to the region of extremely \textit{short-wavelength} fluctuations ($l_f < b$) and small deterministic correlation lengths ($l_C < 10^{-3}b$). This straight-chain phase is labelled as Phase I (see Fig. \ref{fig_phase}(b) top left). 
Next, for regions $\tilde\Lambda \lesssim 10^{0}$, we find a disordered phase with effectively zero positional order in the chain. Here we find that $l_f> b$, thus we conclude that \textit{long-wavelength} fluctuations of greater than the monomer size are not supported by our colloidal chain. This disordered phase is labelled as Phase II (Fig. \ref{fig_phase}(b) top right). 
Further, focusing in the regime $\tilde\Lambda > 10^{1}$ and $\Lambda < 10^{-3}$ corresponds to the case where deterministic rotations are instantaneous across the system ($\tau_f << \tau$), such that the chemical interactions are instantaneously correlated. Here one has that $l_C \gtrsim 10^{3}b$, thus correlations much larger than the system size. The dynamics of the chain displays ``frustrated" behavior, not attaining any non-trivial spatial structure as there is not enough time to respond to instantaneous chemical gradients.
% We note that such a phase was not reported in previous models incorporatin \cite{subramaniam2024rigid} due to the existence of chemical trails that break the forward-backward symmetry; this is also discussed in Appendix \ref{sec:hist_dep}. 
This frustrated phase is labelled as Phase III (Fig. \ref{fig_phase}(c) bottom left).
The remaining region corresponds to that of the C-shape (Phase IV). We thus conclude that the C-shape corresponds to regimes where the correlation lengths $l_C$ are of the order of the chain length, which is in addition stable to short-wavelength fluctuations (of typical distance less that one monomer radius). It is this dynamical steady-state (DSS) phase that we study in the rest of this paper.\\

We also note that, within the C-shape phase, its existence as a DSS depends on the choice of initial conditions of Eq. (\ref{eq:init_cond}). Deviating from these does not render a universality of the DSS, unless one includes the existence of chemical trails in the model \cite{kumar2023emergent, subramaniam2024rigid}. This scenario is further discussed in Appendix \ref{sec:hist_dep}. If one deviates from the purely chemo-repulsive scenario considered here, the presence of trails is also necessary for other non-trivial DSSs, such as a swimming/undulating state. However, various \textit{structural} SSs (e.g. designer crystallites that break handedness symmetry) are possible via the instantaneous chemical deposition model considered here, though one would have to significantly alter from the fully chemo-repulsive scenario, and incorporate mixed (anti-symmetric and/or non-reciprocal) interactions \cite{subramaniam2024rigid}. 

\subsection{Characteristics of the C-shape}
\label{sec:condition_Cshape}
Beyond the (log) curvature of the chain, it is instructive to further distinguish the C-shape from other possible DSSs that can be attained from simulation. These DSSs correspond to those from the phase diagram in Fig, \ref{fig_phase}, but in addition those that are (a) in the same $(\Lambda, \tilde{\Lambda})$ coordinate but of different $N$, (b) adding rotational noise \cite{subramaniam2024rigid, golestanian2009anomalous}, and/or (c) chemical trails in the dynamics. Fluctuations about other DSSs, could in principle be legitimately computed as well. For instance in the Phase I of $\Lambda \rightarrow \infty$ these would trivially reproduce an early-time growth for the height fluctuations, similar to that for a driven EW model, as shown explicitly in Appendix \ref{sec:prob_dist}. We emphasize again that the fluctuations reported in this work correspond to those exclusively about the C-shape. To emphasize the uniqueness of this shape let us list some generic properties of this DSS: 
\begin{enumerate}
    \item Continuously varying orientations along the chain
    \item Finite {mean} curvature of the chain %(or the averaged local curvature)
    \item Ballistically propelling in one direction
    \item Positional and orientational symmetry along the body(y) axis
\end{enumerate}

As explained in Appendix \ref{sec:app_dss}, there are various other DSSs that meet some of the above requirements but not all. 
 For instance the aforementioned Phase I DSS satisfies only (i), (iii) and (iv). The C-shape is unique in that it is the only DSS that meets requirements (i)-(iv), with its propulsion perpendicular to the direction of the body axis. It is in this specific phase in which the fluctuations are computed. 

%%%==============================
%%%==============================
%%%==============================
%%%==============================
%%%==============================
%%%==============================
\begin{table*} [t!]
    \centering
    \resizebox{0.96\textwidth}{!}{%
  \begin{tabular}{|c|c|c|c|c|c |c|c|}
\hline
Section& Quantity & $\beta^{(1)}$ & $\beta^{(2)}$ & $\alpha$ & $z=\alpha/\beta$   & Figure & Remarks \\
\hline
\hline
{ \ref{sec:VA}} & $W_h$ & $\, \,0.25\pm0.01\, \,$ & $\, \,1.73\pm0.01$ \, \,& $\, \, 0.90\pm0.03\, \,$ & $\, \,0.52\pm0.02\, \,$   &  \ref{fig3} &Phase IV - positional  \\
\cline{1-8}
 \ref{sec:VB}& $W_{\theta}$ & $\, \,1.00\pm0.01\, \,$ & NA & $\, \,-0.48\pm0.02\, \,$ & $\, \,-0.48\pm0.02\, \,$  &  \ref{fig4} & Phase IV - orientational \\
\hline
\multirow{2}{*}{ \ref{sec:VC}} & $W_h$ & $\, \,0.25\pm0.02\, \,$ & $\, \,1.73\pm0.03\, \,$ & $\, \,0.81\pm0.04\, \,$ & $\, \,0.47\pm0.03\, \,$  & \multirow{2}{*}{\ref{fig_app_drpos}} &  \multirow{2}{*}{Finite $\tau_r$} \\
\cline{2-6}
& $W_{\theta}$ & $\, \,1.00\pm0.02\, \,$ & NA & $\, \,-0.48\pm0.02\, \,$ & $\, \,-0.48\pm0.02\, \,$   &  &  \\
\hline
\multirow{1}{*}{\ref{sec:locally_flat}} & $W_h$ & $\, \,0.25\pm0.02\, \,$ 
& NA 
& $\, \, 1.00\pm0.01\, \,$ & $\, \,4.00\pm0.30\, \,$   &  \ref{fig5} & Locally flat - positional   \\
 % \hline
 \hline
 Appendix \ref{sec:app_wtheta_flat}
 & $W_{\theta}$ & $\, \,1.00\pm0.01\, \,$ & NA & $\, \,1.02\pm0.02\, \,$ & $\, \,1.02\pm0.02\, \,$  &  \ref{fig8ThetaF} &   Locally flat - orientational   \\
 \hline
 % \multirow{2}{*}{Appendix \ref{sec:hist_dep}} & $W_h$ & $\, \,1.41\pm0.02\, \,$ & NA & NA & NA  & \multirow{2}{*}{\ref{fig10History}} &  \multirow{2}{*}{With chemical trails} \\
 % \cline{2-6}
 %  & $W_{\theta}$ & $\, \,2.11\pm0.02\, \,$ & NA & NA & NA  &  &  \\
 %  % \hline
 %  \cline{1-8}
\end{tabular}
    }
    \caption{\label{tab:exp}Table of exponents defined in equations (\ref{eq:d2h}-\ref{eq:theta_scaling}) of section \ref{sec:chemo-rep}. Note that $z$ is defined such that $\beta^{(2)}$ is in the denominator for the first and third rows, whereas $\beta^{(1)}$ is in the denominator for the others (orientational and locally flat exponents). For the last two columns, $z$ is taken to be $z'$ implicitly. 
    %For comparison, we note that for a one-dimensional interface: 
    % $\alpha^{\mathrm{EW}} =1/ 2$, $\beta^{\mathrm{EW}} =1/ 4$; 
    % $\alpha^{\mathrm{KPZ}} =1/ 2$, $\beta^{\mathrm{KPZ}} =1/3$ \cite{edwards1982surface, kardar1986dynamic}.
    }
\end{table*}
%%%==============================
%%%==============================
%%%==============================
%%%==============================

\subsection{C-shape selection}
In Phase IV, as mentioned previously, the C-shape is always attained under the initial conditions of \ref{eq:init_cond}. We note that in such an autophoretic chain with chemical trails this DSS is the universal solution regardless of initial conditions \cite{kumar2023emergent, subramaniam2024rigid} (see Appendix \ref{sec:hist_dep})). However, it is instructive to ask to what extent the C-shape is attained under different initial conditions. \\
To this end, we apply a hysteretic protocol by varying $\Lambda$ and carrying on the final state of a given simulation to the initial conditions of the simulation at the next time step. The order parameter used to identify this DSS is the absolute value of the curvature, $| \kappa |$ defined in Eq.~\eqref{curv_def}. %The results of these are presented in Fig. \ref{fig_phase}(c). 
Two separate cases are presented in Fig. \ref{fig_phase}(c), corresponding to decrease of $\Lambda$ starting from Phase I (the stiff chain; Fig. \ref{fig_phase}(c), \textit{top}), and increase of $\Lambda$ starting from Phase IV (the C-shape; Fig. \ref{fig_phase}(c), \textit{bottom}). We observe from the former that the stiff-to-C transition is irreversible. Thus, a C-shape DSS cannot lose its spatial curvature via a decrease in self-phoretic torques. From the latter we see that the C-to-stiff transition is hence not permitted; both forward and backward sweeps of the protocol do not change the overall curvature. We also note that we have separately varied $\tilde{\Lambda}$ to and from Phase IV (not shown here); the results of these indicate that selected random configurations of Phase II may indeed transition to IV, but again there is no clear reliable reversibility of transition to define a hysteresis loop.
This irreversibility of the first transition implies that such a hysteresis curve and hence a mapping to known classes of equilibrium or dynamical phase transitions is not well defined (indeed one may have to devise an alternate order parameter); we leave the exact characterization of the nature of the transition to and from the C-shape for future work. \\

We merely conclude this section by emphasizing that the C-shape is the attractor (global minimum) of our EOM (\ref{eq:dyn}) with initial conditions of (\ref{eq:init_cond}).
The subsequent results of this work are thus based on the assumption of instantaneous chemical depositions and the aforementioned initial conditions; we note that these may change upon incorporation of chemical trails. We display in Appendix \ref{sec:hist_dep} that identical growth exponents are obtained when chemical trails are present; where the C-shape is instead a universal attractor (independent of initial conditions) \cite{subramaniam2024rigid}. \\

\section{Interface fluctuations}\label{sec:chemo-rep}
%%------------------------
% \begin{figure*}[t]
%     \includegraphics[width=0.8\textwidth]{fig02.png}
%     \caption{Schematic of dynamical evolution of colloidal chain. From left to right: evolution of the chain at different times $t=0$, until the final $t=T$. At times above $\tau$ and $\tau_f$ the chain acquires the steady-state C-shape. Note that the chain propels in the positive $x$ direction, and that the time points are not linearly scaled. The inset on each show the monomer orientations for a particular segment at each time. $N'$is labelled in red (middle panel), along with $\Delta h$ (rightmost panel) and $\theta$ (in inset). }
%     \label{fig0}
% \end{figure*}

For the C-shape chain propelling in the positive $x$-direction, one can define a height function $h(y)$ along the chain. In particular $h$ corresponds to the $x$-displacement of each monomer from the mean height. An example of its temporal evolution is shown in Fig. \ref{fig_phase}(d). One can thus choose a particular length segment along the chain and compute the height fluctuations, and measure its root-mean-squared value
\begin{align}
    W_{h}(t) = \sqrt{ \langle (\Delta h)^{2}(t) \rangle } = \sqrt{\langle (h_{i}(t) - \langle h(t) \rangle)^{2} \rangle }  \, ,
\label{eq:d2h}
\end{align}
for the C-shape. Here, $\langle \rangle$ is taken over both the segment and realizations. We can do the same for the angular fluctuations, and define 
\begin{align}
    W_{\theta}(t) = \sqrt{ \langle (\Delta \theta)^{2} (t)\rangle } = \sqrt{ \langle (\theta_{i}(t) - \langle \theta(t) \rangle)^{2} } \, .
\label{eq:d2th}
\end{align}
The quantities of interest are both the dynamical ($W_h(t \ll \infty)$  and $W_\theta(t \ll \infty)$) and steady-state ($W^{\mathrm{ss}}_h(t \rightarrow \infty)$ and $W^{\mathrm{ss}}_\theta(t \rightarrow \infty)$) properties of these fluctuations. These are:
\begin{subequations}
\begin{align}
    W_h(t) & \sim t^{\beta_h},
    \qquad W^{\mathrm{ss}}_h  \sim N^{\alpha_h},
%    \qquad W^{\mathrm{ss}}_h  \sim   \left(N'\right)^{\alpha'_h},
    \\
    W_\theta(t) & \sim t^{\beta_\theta},\qquad W^{\mathrm{ss}}_\theta  \sim 
    N^{\alpha_\theta}.
    %\qquad
    %\textcolor{blue}{W^{\mathrm{ss}}_\theta \sim ??}.   
\end{align}    
\end{subequations}
Here, $(\beta_h, \beta_\theta)$ and
$(\alpha_h, \alpha_\theta)$ are sets of growth and roughness exponents, respectively. As we show below, the positional fluctuations can be expressed via the FV scaling law \cite{vicsek1984dynamic}
\begin{align}
    W_h & \sim  N^{\alpha_h} \, 
    f_h(t/N^{z_h}),\qquad z_h = 
    %\frac
    {\alpha_h}/{\beta_h},
\label{eq:fv_scaling}
\end{align}
where the scaling function $f_h$ is given as:
\begin{align}
    f_{h}(t/N^{z_h}) \propto \begin{cases}
    t^{\beta_{h}} &  t << t_h^{*}(N) \\
    1 &  t >> t_{h}^{*}(N) \\
\end{cases}
\label{eq:scaling_h_limits}
\end{align}
For the orientational sector, we find that, though there is a growth followed by a plateau, there is a \textit{persistent smoothening} phenomena, where there is no system-size dependent saturation time scale, but instead a system-size dependent scaling of the fluctuations at \textit{all times}, which is instead negative (hence smoothness). This is further elaborated on below. In this case, the scaling reads:
\begin{align}
    W_{\theta} & \sim  N^{\alpha_{\theta}} f_{\theta}(t).
\label{eq:theta_scaling}
\end{align}
Here, we have defined:
%and
\begin{align}
    f_{\theta}(t) \propto \begin{cases}
    t^{\beta_{\theta}} &  t << t_{\theta}^{*} \\
    1 &  t >> t_{\theta}^{*} \\
\end{cases}
\label{eq:scaling_th_limits}
\end{align}
with the important difference between the two being the $N$ dependence of the characteristic saturation time of fluctuations $t^{*}$, as explained below. We note that given (\ref{eq:scaling_th_limits}), $W_{\theta}^{\text{SS}}$ can be equivalently be taken to be the time average over the entire simulation. \\

%%%%============
%%%%============
\begin{figure*}
    \centering
    \includegraphics[width=0.68\textwidth]{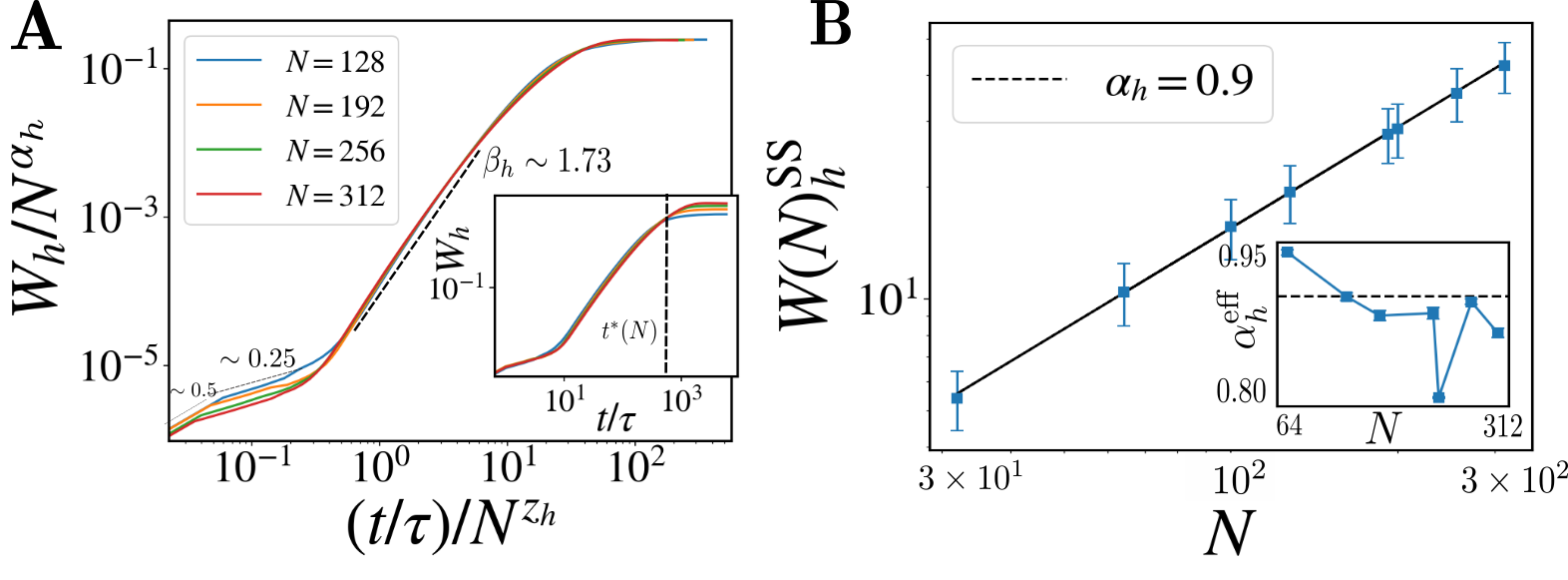}\\
    \caption{Scaling of height fluctuations. \textbf{A} $W_h$ for different chain lengths $N$, with the three exponents $\beta_h$ indicated via dashed black line, fit within $(t/\tau N^{z_h}) \in [1, 20 ]$. The same scaling is plotted in the \textit{inset}, with $t_{h}^{*}(N)$ labelled in dashed vertical line.  \textbf{B} Roughness scaling for $W_h^{\text{SS}}(N)$, with effective exponent in inset. Note that for the effective exponents $N \in [ 64, 312]$ (excluding the smallest $N=32$, due to forward-difference) Here, $\tau_f = 0.2$, $\Lambda=0.2$, $\tilde{\Lambda}=0.01$.}
    \label{fig3}
\end{figure*}

To compute the quantities $W_h(t)$ and $W_\theta(t)$ [Eq. (\ref{eq:d2h}) and (\ref{eq:d2th})] respectively, we discuss two possibilities. There are two separate segments about which fluctuations are measurable. Let us label the length of this segment to be $N'$. The first obvious choice of the segment is $N'=N$, corresponding to the average over the entire C-shape. The second corresponds to that of a \textit{locally flat} region; here $N'<<N$, which is taken to be sufficiently far from the edges of the chain. 
Though established results exist for fluctuations about curved segments of an interface \cite{barabasi1995fractal}, these typically correspond to that of a (subset of a) \textit{circular} interface, thus different from that of the C-shape (see Appendix \ref{sec:app_curvature}).
For the latter, it is to be noted that fluctuations of growing (circular) interfaces are typically measured in this regime 
\cite{takeuchi2010universal, takeuchi2012evidence} ; the scaling properties in this region are almost always equal to that of globally measured quantities - for instance in \cite{kim1991surface} for a flat surface and \cite{takeuchi2010universal} for a curved surface. We discuss the specific instances of equivalences and/or differences for the C-shape below. \\

All results presented in this paper are computed for values of $N \in [32,64,100,128,192,200,256,312 ]$, with curves for specific $N$ displayed (and labeled) in each respective figure. The results for the incorporation of rotational noise are computed for $N \in [64,128,192,256,312 ]$. The choices for $N'$ are explicitly stated in the respective sections.

\section{Results}

A summary of the results of this paper is presented in Table \ref{tab:exp}, wherein the respective Figures are referenced. It is to be noted that the value of $\beta$ for each respective figure is obtained through the best  possible collapse of the scaling relations \ref{eq:fv_scaling} and \ref{eq:theta_scaling} (by eye). The values for $\alpha$ on the other hand are obtained by a least-squares fit on the steady state values of $W$ for each system size. For each plot of $W$, the fitting windows used for the in-figure $\beta$ line of best fit is mentioned in the caption.\\

% \begin{table}[h!]
% \centering
% \begin{tabular}{c|c c|c}
% \multirow{2}{*}{\multicolumn{1}{c|}{Left block}} 
%     & A1 & B1 & \multirow{2}{*}{\multicolumn{1}{c}{Right block}} \\ 
%     & A2 & B2 & \\ 
% \hline
% X   & 1  & 2  & Y \\
% Z   & 3  & 4  & W \\
% \end{tabular}
% \end{table}
\begin{figure*}[ht!]
    \centering
    \includegraphics[width=0.645\textwidth]{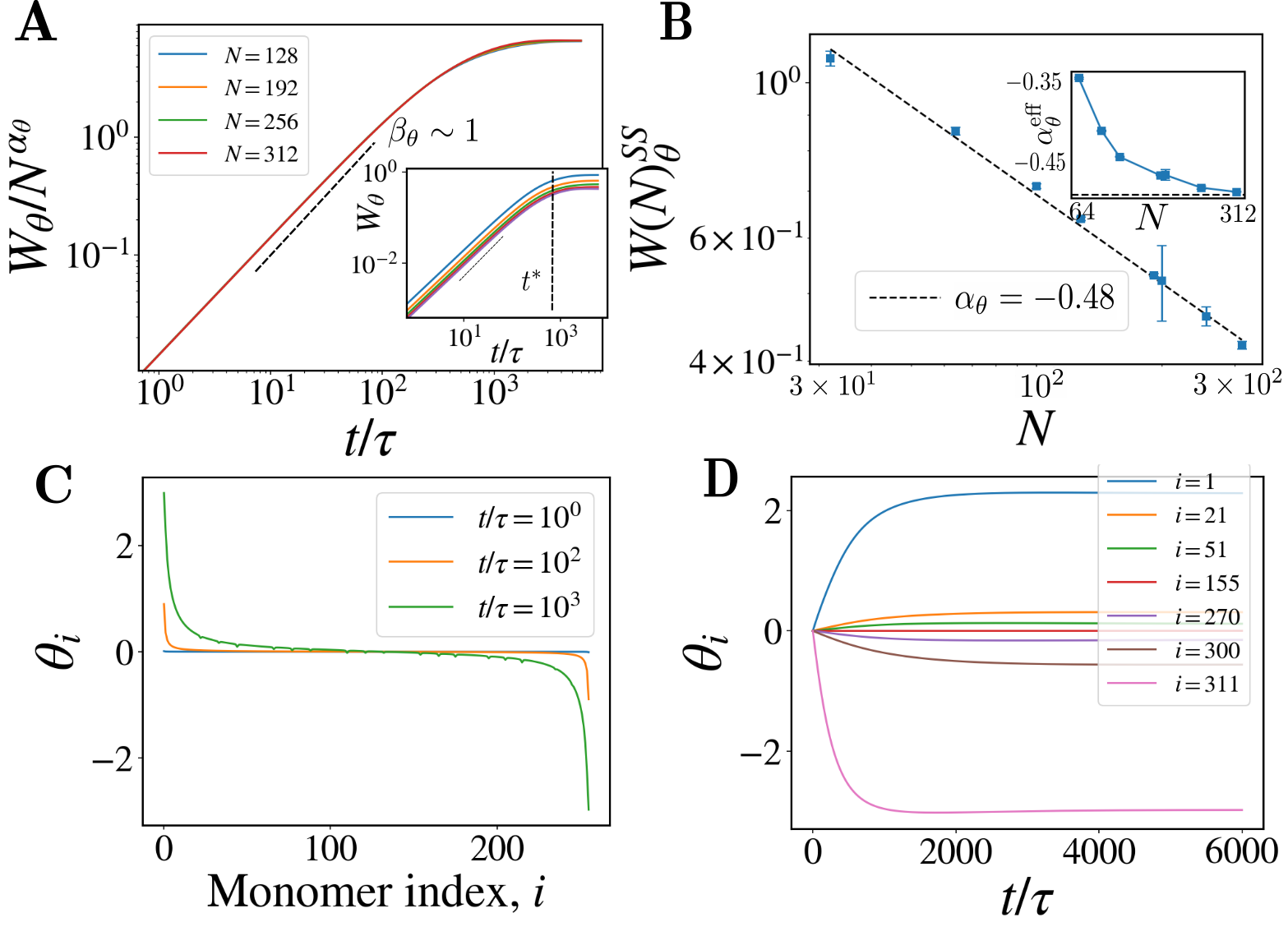}
    \caption{Scaling of the orientation fluctuations. \textbf{A} $W_{\theta}$ for different chain lengths $N$, with the dynamic $\beta_{\theta}$ labelled (fit between $(t/\tau N^{z_h}) \in [10, 200 ]$). The same scaling is plotted in the \textit{inset}, with $t^{*}$ labelled in dashed vertical line. \textbf{B} (Anti-) Roughness scaling for $W_{\theta}^{\text{SS}}(N)$, with effective exponent in the inset. \textbf{C} Orientation profile of the C-shape, taken at different time points. \textbf{D} Dynamical evolution of $\theta_i$ for selected monomers $i$. Here, $\tau_f = 0.2$, $\Lambda=0.2$, $\tilde{\Lambda}=0.01$, and $N=311$ (for panels C and D).}
    \label{fig4}
\end{figure*}

\subsection{Height fluctuations}\label{sec:VA}
The results for the interface height fluctuations $W_h$ are shown in Fig. \ref{fig3}(A). We see that there are four separate dynamical regimes, with three growth exponents which are defined as. 
\begin{align}
W_h(t) \sim        \begin{cases}
          t^{\beta^{(0)}_h},\qquad 
        t\ll \tau,t\ll \tau_f,\\
          t^{\beta^{(1)}_h},\qquad \tau \ll t<\tau_f,\\
         t^{\beta^{(2)}_h},\qquad  t>\tau_f.\\
    \end{cases}
\end{align}
In all the three cases above $t<t_h^*$. Around $t\sim t_h^*$, the fluctuation $W_h$ saturates 
to its steady-sate value $W_h^{\mathrm{ss}}$ due to the finite system size. We now discuss these four regimes in detail.
\begin{enumerate}[(i)]
    \item \textit{Early-time regime I}. 
This corresponds to times $t<<\tau$ (first few time steps of simulation). 
Here the growth of the interface height is dominated purely by noise, the mean-squared displacement of the height scales as $t$, and thus $\beta^{(0)}_h \sim 0.5$.
\item \textit{Early-time regime II}. Now the colloids start to feel interactions from their respective neighbors which leads to slowing down of the pure diffusive growth. 
Deterministic chemical self-interactions, though, are yet to set in. Here, $\beta^{(1)}_h  \approx 0.25$. These first two regimes correspond to those observed in the EW model \cite{edwards1982surface}. For $t < \tau_f$, it is straightforward to show that our chain approximates an EW interface, as shown in  Appendix \ref{sec:prob_dist}. These early time scalings correspond to (as expected) a purely diffusive regime, equivalent to random deposition of particles on a substrate accompanied by surface diffusion. An alternate way to characterize the early time growth is via the probability distribution of $W^{2}_h$; this (and a comparison with the corresponding EW distribution) is presented in Appendix \ref{sec:prob_dist}.
\item \textit{Super-ballistic} regime: In this regime $t > \tau_f$ with $t<t_h^*$.  During this time, the self-phoretic interactions cause the chain to morph into a C-shape topology. As a result, $W_h$ swells up to its steady-state value, with a super-ballistic exponent ($\beta_h^{(2)}$, quoted below). Let us study the behaviour of these fluctuations by defining a correlation length $l_c$. Via dimensional analysis, $l_{c} \sim b (\frac{t}{\tau})^{\frac{1}{z_h}}$.
In this region $l_c$ typically corresponds to a finite fraction along the interface. For instance, at $t=10\tau$ time steps into the simulation (at the onset of swelling, see inset of Fig. \ref{fig3}(A)), $l_c \approx 53b$. These correlations further build-up until the next regime is reached. 
\item \textit{Steady-state} regime. This regime corresponds to $t>>\tau_f$ and $t >> t_h^{*}$. We identify $t_h^{*} = N^{z_h} \tau$, a system-size dependent characteristic timescale over which the fluctuations reach a steady-state (see inset of Fig. \ref{fig3})(A). For example for $N=256$ (Fig.\ref{fig3}(A) \textit{inset}, green), $t_h^{*}\approx 27.6 \tau \sim 10^{2} \tau_f$. Here, the fluctuations cease to grow
and $W_h$ reaches a plateau. The aforementioned correlations now scale as $(\frac{t}{\tau})^{\frac{1}{z_h}} \sim N$ (spanning the entire chain length). \textit{System size scaling}. Fig. \ref{fig3}(B) shows the scaling of $W_h^{SS}$ with the chain length $N$; a roughness exponent of $\alpha_h \approx 0.9$ is found. The errorbars on  coordinate points are computed by taking the standard deviation of $W_h$ over 50 realizations in the steady-state. In the inset the effective exponent $\alpha_h^{\text{eff}} = \frac{d \log W_h^{SS}}{d \log N}$ is plotted. A moderate convergence is observed to the best fitted exponent of $0.9$ (black dotted horizontal line in inset). 
\end{enumerate}
% (i)  \\
% (ii)  \\ 
% (iii) \\
% (iv) \\

% \\

The growth and roughness exponents, respectively characterizing the height fluctuations in regimes (iii) and (iv)  are: 
\begin{align}
    \beta^{(2)}_h = 1.73 \pm0.01 \label{eq:exp_cshape1}\\
    \alpha_h = 0.90 \pm 0.03 \label{eq:exp_cshape2}
\end{align}
From \eqref{eq:exp_cshape1}, \eqref{eq:exp_cshape2} and Fig.~\ref{fig3},  we find that the regimes (iii) and (iv) exhibit a clear FV scaling with a roughness exponent $\alpha_h$ and a dynamic exponent $z_h=\frac{\alpha_h}{\beta_h} \approx 0.52$. 
To the best of our knowledge, this set of exponents does not correspond to any previously reported universality class.
We thus interpret this as a novel ``C-shape roughness" scaling, distinct to those found for either circular or flat interfaces \cite{takeuchi2010universal, family1985scaling}. The super-ballistic exponent can be appreciated by noting that as the C-shape formation takes place, displacements of each monomer from the mean is enhanced in either direction (along the x-axis) along the chain. These squared deviations from the mean height are in addition symmetric along the y-axis. $l_c$ is clearly enhanced due to this symmetry. It is to be noted that in a conventional ballistic scaling one simply has uni-directional deviations along the propulsion axis; in our system this is applicable to the mean $h(y)$; clearly $l_c$ in this case is smaller than that for a symmetric C-shape. We suggest that these combined effects give rise to the super-ballistic growth exponent $\beta_h$. The emergence of a super-ballistic exponent can be rationalized via a linearised analysis of the EOM \ref{eq:dyn}. This is presented in Appendix \ref{sec:app_super_ball}. It is to be noted that varying $\tau_r$ changes the time at which super-ballistic growth sets in (exponent and roughness is unchanged). The roughness exponent $\alpha_h$, on the other hand, indicates the relative growth of height fluctuations when $l_c$ spans the system size. Here, the system is at $\sim 10^{2} \tau_f$ and the C-shape is fully formed. We note that the value we report for $\alpha_h \approx 0.9$ is greater than  the standard roughness exponent encountered in dynamics without height conservation ($\alpha_h \approx 0.5$)~\cite{edwards1982surface, kardar1986dynamic} and smaller than that in models with height conservation~($\alpha_h \approx 1.5$)\cite{ sarma1991new}.
% \textcolor{black}{The citations 38,39 of $\alpha_h=1.5$ are for 2d interfaces. Maybe it is enough to cite Das Sarma here since it corresponds to 1d interface.}
%In addition it is larger than most numerical deposition models such as for instance the solid-on-solid growth model \cite{kim1991surface}. 
We elaborate on these findings further in Sec.~\ref{sec:summ}. 
%It is however, smaller than that to be expected of a diffusive height of conserved mass \cite{sarma1991new}.
%, where the exponent is $2$ \cite{sarma1991new} 
% (in one-dimension). 
%Th%is is to be expected, since as curvature sets i%n, local mass is no longer conserved along the chain. \textcolor{black}{The last statement is not very clear to me. Also, the exponent $\alpha$ reported in \cite{sarma1991new} is 1.5, not 2.}\\ 

%%%==============
%%%==============
\begin{figure}[t]
    \centering
    \includegraphics[width=0.485\textwidth]{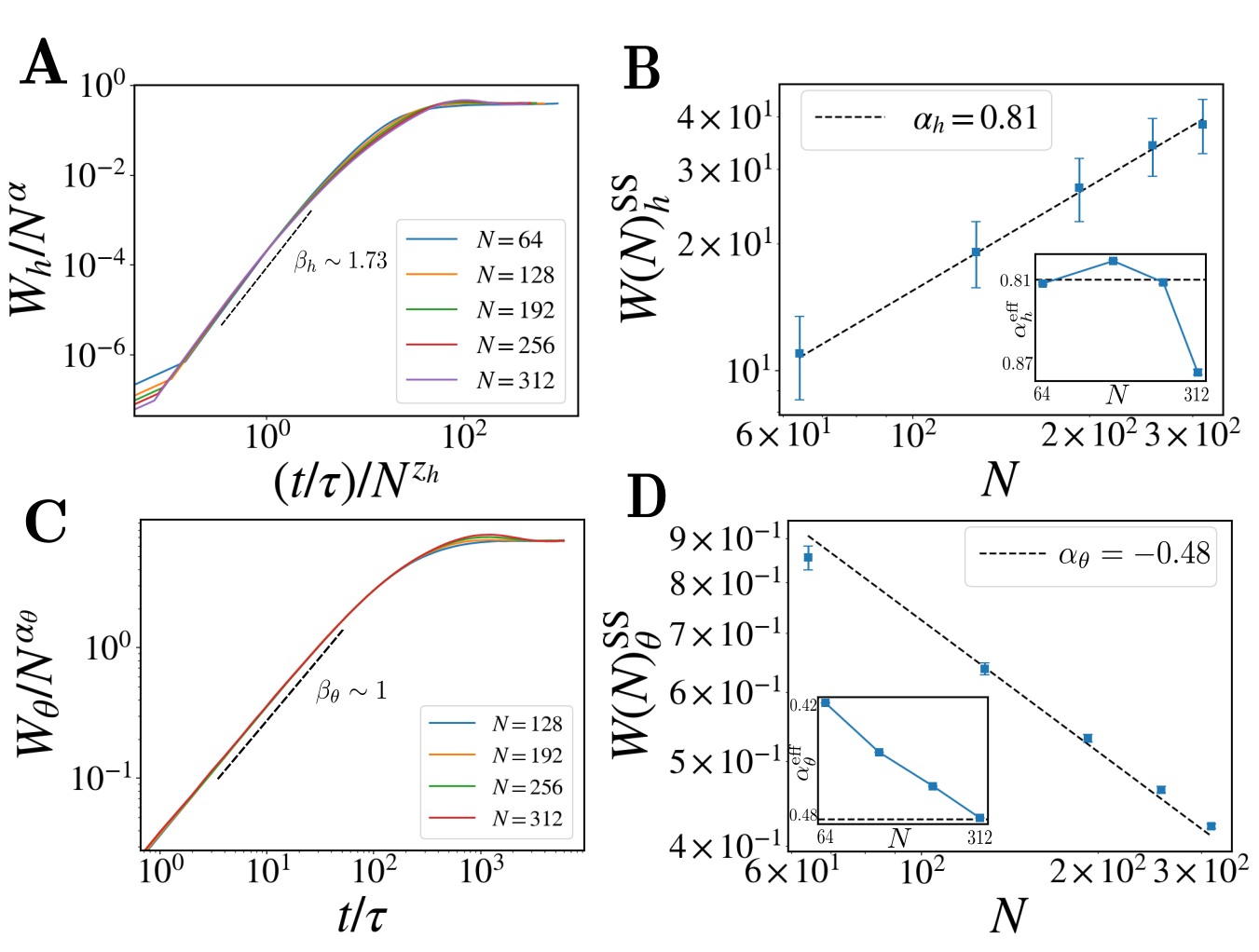}
    \caption{Height and orientational fluctuations for finite $\tau_r$. \textbf{A} $W_{h}$ for different chain lengths $N$, with the dynamic $\beta_{h}$ labelled (fitting window is $ t/(\tau N^{z_h}) \in [0.5,5 ]$). Scaling for $W_h^{\text{SS}}(N)$, with effective exponent in the inset. \textbf{C} $W_{\theta}$ for different chain lengths $N$, with the dynamic $\beta_{\theta}$ labelled (fitting window is $ t/(\tau N^{z_{\theta}}) \in [4,80 ]$). \textbf{D} (Anti-) Roughness scaling for $W_{\theta}^{\text{SS}}(N)$, with effective exponent in the inset. Here, $\tau_f = 0.2$, $\Lambda=0.2$, $\tilde{\Lambda}=0.01$ and $\tau_r = 70$.}
\label{fig_app_drpos}
\end{figure}
%%%==============
%%%==============

%%======================================
%%======================================
\subsection{Orientational fluctuations}
\label{sec:VB}
%%======================================
%%======================================
 The orientational fluctuations are displayed in Fig. \ref{fig4}(A) (and inset). There are two regimes displayed:
(i) \textit{Ballistic} regime. This corresponds to deterministic orientational changes as the C-shape forms.
  (ii) \textit{Steady-state} regime. $W_{\theta}$ is distinguished from $W_{h}$ in that there is no $N$ dependent $t^{*}$ during which the fluctuations relax. A rough estimate of this from Fig. \ref{fig4}(A) would be $t_{\theta}^{*} \sim 50\tau_f$, thus of the same order as $t_{h}^{*}(N)$. Further, $l_{c}$ spans the system size at \textit{all times} as displayed in Fig. \ref{fig4}(A) \textit{inset}.\\

The ballistic regime arises due to the purely deterministic contribution to the orientational changes that sets in during $t  \sim 5\tau_f \approx \tau$. The emergence of the ballistic regime can be appreciated by consideration of the orientation profile $\theta_i$ in Fig \ref{fig4}(C) for $N=312$. We see that at all times $\theta_i$ maintains strong correlations across the chain, with $\theta_{i} = -\theta_{N-i}$ (anti-symmetry; here $i$ is the monomer index). 
Such an
%bidirectional and 
anti-symmetric profile of the orientation field along the y-axis, along with the dynamics 
%(as opposed to a ballistic positional exponent arising from uni-directional displacements), with the mean-squared thus giving a net positive growth. Such a symmetrical dynamical evolution 
thus enforces $l_{c}$ to span the entire chain length. 
Further, as the steady-state is approached, a large number of orientations remain close to their initial value (\ref{fig4}(d)). For instance, if one takes a threshold $\theta^{*}=0.2$ (corresponding to $|\frac{\max(\theta) - \theta^{*}}{\max(\theta)}| \approx 0.92$ deviation from the edge monomer angles), we find that a fraction of $\approx .89$ of monomers on the chain remain roughly equal to their initial value, with the remaining $\approx .11 $ drastically varying at the edges (``orientational stiffness"\cite{kumar2023emergent}). 
%\textcolor{blue}{We also note that we keep the orientation noise $D_r$ small such that the chain evolves only to the C-shape topology, and not to other states mentioned in the phase diagram [Fig.~\ref{fig_phase}].} 
We note that here $\tau_r \rightarrow \infty$; the results for finite $\tau_r$ are presented in Appendix \ref{sec:VC}; there is in addition an early time diffusive regime, though that does not affect the results presented here 
\footnote{As explained in \cite{subramaniam2024rigid}, 
sufficiently large rotational noise prohibits the C-shape formation. Thus, 
``positive $D_r$'' wherever mentioned here implicitly assumes $\frac{\tau_f}{\tau_r} = \frac{b^{3}D_r}{\chi_r} << 1$. 
Here, $\frac{\tau_f}{\tau_r} = 0.01$.}.\\

We further report that $W_{\theta}^{SS}$ scales \textit{negatively} with the chain length [Fig.~\ref{fig4}(B)]. We may thus call this a \textit{smoothness} exponent. This form of orientational stabilization with the system size, as opposed to conventional ``roughness", and implies a suppression of long-ranged orientational fluctuations at larger system sizes. The source of emergence for such a smoothening exponent is the increased orientational rigidity along the chain, which increases with chain length \cite{kumar2023emergent}. The angular distribution along the C-shape becomes increasingly uniform, thereby making the chain stiffer and more polarized along the x-axis. The RMS deviations from the the mean orientation thus decreases with system size (a larger fraction of the chain is polarized along the x-axis). This is in contrast to the positional steady-state fluctuations, where the mean $h(y)$ varies with chain length, and fluctuations about this exhibits roughness.

%%%======================================
%%%======================================
%%%======================================
\subsection{With finite rotational noise}
\label{sec:VC}
%%%======================================
%%%======================================
%%%======================================
The exponents discussed so far correspond to the case $D_r=0$ (equivalently $\tau_r \rightarrow \infty$) in \eqref{eq:dyn}.
 Incorporating finite rotational noise $D_r$ may indeed change the possible DSS obtained from simulations. We do not report those here; but simply mention that $\tau_r \approx 10$ sets a threshold above which the C-shape is the attractor. In this section, we further investigate whether the aforementioned findings carry over to the case of finite $\tau_r$. Indeed, we find that $\tau_r \gtrsim 20$ an identical FV scaling for $W_h$ is obtained, whereas $W_{\theta}$ only approximates the scaling law and collapse (\ref{eq:scaling_th_limits}) - See Fig. \ref{fig_app_drpos}. The effect of randomness prohibits a system-spanning $l_c$. Nevertheless, a smoothness exponent with an identical scaling can be obtained. 

For $\tau_r \lesssim 10$ there is no longer a propelling C-shape, 
and indeed other DSSs are observed. 
The reason for this is that the C-shape DSS is sensitive to long-wavelength fluctuations in this regime, such that any fluctuations on the length scale $\frac{b^{2}D_r}{v_s}$ is greater than typical monomer lengths. 
It is to be noted that for equilibrated systems $\tau_r = \frac{8\pi \eta b^{3}}{k_b T}$, 
with $\tau_t  = \frac{6 \pi \eta b^{2}}{k_b T}$ ($\eta$ is 
the viscosity of the medium). 
Thus the effect of rotational noise and hence potential loss of the 
C-shape DSS may set in for smaller $b$. 
This trade-off and its potential experimental relevance 
for the system considered here is discussed further in Sec.~\ref{sec:summ}.

%%%===============
%%%===============
%%%===============
\begin{figure}[b!]
    \centering
    \includegraphics[width=0.475\textwidth]{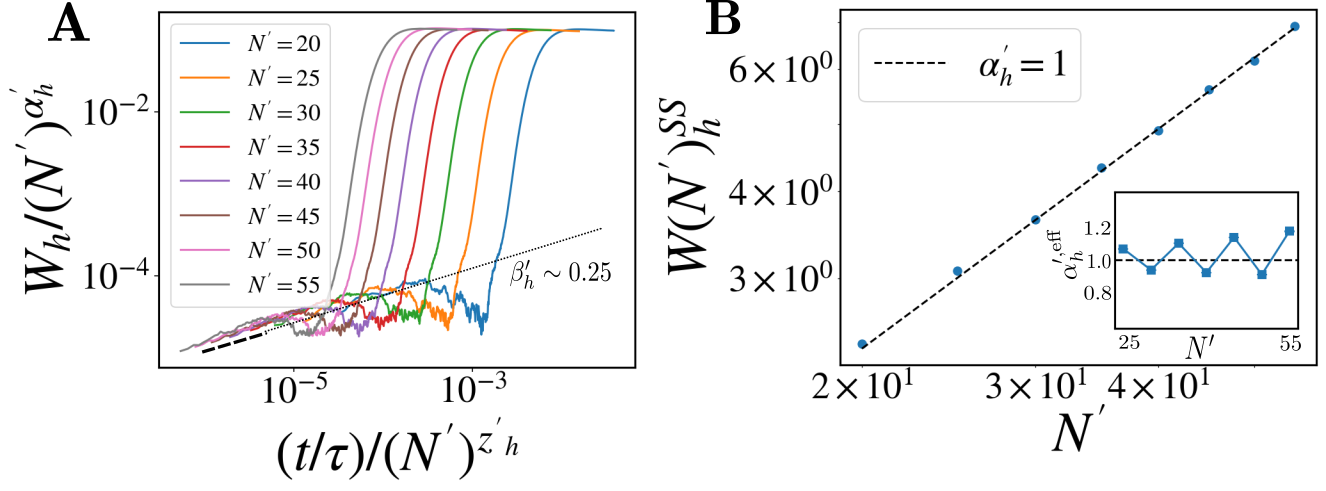}
    \caption{Scaling of height fluctuations in locally flat region. \textbf{A} $W_{h}$ for different locally flat lengths $N'$, with the dynamic $\beta_{h}'\approx 0.25$ indicated via dashed black line, fit between $(t/\tau N'^{z_{h}'}) \in [5\times 10^{-6}, 10^{-5}]$. \textbf{B} Roughness scaling for $W_{h}^{\text{SS}}(N')$. Here, $N=312$ and $\tau_f = 0.2$, $\Lambda=0.2$, $\tilde{\Lambda}=0.01$.}
    \label{fig5}
\end{figure}
%%%===============
%%%===============
%%%===============

\subsection{Locally flat regimes}
\label{sec:locally_flat}
In the study of fluctuations of circular interfaces, the fluctuations are typically measured about a locally flat region along the growing front \cite{takeuchi2010universal, takeuchi2012evidence, barabasi1995fractal}; and the roughness scalings thus correspond to those of a locally flat region. We repeat the same procedure here, and vary $N'$ under the constraint of $N'<<N$. In this section, $\frac{N'}{N}<0.18$ such that the interface is effectively flat. The results are presented in Fig. \ref{fig5}. Probing the locally flat regime, we find that the height fluctuations in the early-time ($\beta^{(1)}$) and steady-state regimes, respectively, can be collapsed onto a FV-type scaling with growth exponent $\beta_h'  \approx 0.25$, %thus equivalent to that of the EW growth exponent \cite{kardar1998nonequilibrium, barabasi1995fractal}. 
and roughness exponent $\alpha_h' \approx 1$. This would imply a dynamic exponent $z_h' =4$. This combination of exponents, interestingly, is reminiscent of interfaces with conservation laws in {\em two dimensions} ~\cite{mullins1957theory, sarma1991new, CKPZ-Grant, CKPZ-Tirtha}. However, it is to be noted that this implies essentially neglecting the super-ballistic regime, where the data does not collapse. 
The $W_{\theta}$ scaling in this regime is presented in Appendix \ref{sec:app_wtheta_flat}.
%%%=============================%%%%%%%%%%%%%%%%%
%%%=============================%%%%%%%%%%%%%%%%%
\section{Summary and Discussions}\label{sec:summ}
%%%=============================%%%%%%%%%%%%%%%%%
%%%=============================%%%%%%%%%%%%%%%%%
Roto-translational coupling has been well-studied for their various implications on dynamical and collective behaviour in active matter physics. The effect of this coupling on scaling laws has been investigated for the mean-squared displacement - i.e enhanced or anomalous diffusion - in various models (see for instance \cite{golestanian2009anomalous, bechinger2016, kurzthaler2017intermediate}). Here, we instead present the effect of roto-translation-induced global \textit{topological} change, and the resultant novel scaling laws for interface fluctuations. Remarkably, we find that the height fluctuations show an FV data collapse, with a new set of dynamic and roughness exponents; with the growth exponent rationalized via a simple linearised analysis of our model equations presented in Sec.~\ref{sec:2}. The orientational fluctuations, in addition, display novel scaling behaviour, notably \textit{smoothening} with system size, which we attribute to the enhanced orientational rigidity of longer chains. \\

% \subsection{Physical basis for observed exponents}
Some of our observed exponents are rationalizable via comparison with those of well-studied models. The roughness exponent $~0.9$ accompanying the height fluctuations  is higher than that in non-conserved models of total height ($\sim 0.5$)~\cite{edwards1982surface, kardar1986dynamic} and lower than that in conserved models ($\sim 1.5$) \cite{sarma1991new}.
% [\textcolor{black}{It's important to CITE correct examples of $\alpha_h=1.5$ in 1d if we want to keep the previous line. If you have in mind linearized MBE, then cite it (Ref. 17).}]. 
These two correspond to the limit of ``monomeric units" exchanging with the bulk (e.g. evaporation) and being distributed across the surface respectively. The latter limit for instance corresponds to the dynamic exponent we report in the locally flat regime (with monomeric redistribution but discounting the C-shape).
% \textcolor{black}{I'm not sure I follow the argument here. We do not lose any monomer from the chain; so there is no evaporation/ deposition from/in the bulk. I don't think such a simple picture holds for our interface. There is really a coupled dynamics at play which makes the interface different from those that are described purely by a single scalar field. } 
We thus see why the C-shape roughness lies in between these two values - the C-shape does distribute monomeric displacements across the body (thus less suppression of large scale fluctuations), but does so whilst maintaining a fixed bending rigidity. 
% \textcolor{black}{it is actually more rough, since we expect the stiff phase to be simply a driven EW interface. Maybe it is enough to say that anti-alignment interactions lend roughness to the interface height.}).
The growth exponent here being super-ballistic is solely a consequence of topological change due to roto-translational coupling; a simple picture as to how roto-translational coupling could give rise to super-ballistic growth is provided in Appendix \ref{sec:app_super_ball}, where an exact solution to a linearized approximate dynamics is presented (a flat chain with lowest-order roto-translational coupling). For the orientational sector, the smoothness can be appreciated by noting that as the system size increases, a proportionately larger fraction of the monomers are polarized along the x-axis, thus orientational fluctuations at larger length scales get more suppressed. The ballistic growth exponent on the other hand arises due to the deterministic effect of orientational changes that take place near the chain edges. \\

The emergence of novel scaling exponents [Table~\ref{tab:exp}], phases and transitions between them [Fig.~\ref{fig_phase}] demand a thorough hydrodynamic analysis. While we defer the proposition and analysis of such a theory to future work, here we briefly outline some ingredients of a requisite continuum picture. Although individually each colloid can self-propel in any direction, the emergent macroscopic chain spontaneously breaks the isotropy by moving in a chosen direction. When the chain is in its stiff phase (I in Fig.~\ref{fig_phase}), all monomers point in the same direction. However, the C-shape phase (IV in Fig.~\ref{fig_phase}) arises as a result of anti-alignment of monomer orientations. At the minimal level, it may then be hypothesized that the one-dimensional chain can be modeled in terms of two dynamically coupled fields: (i) the local interface height ($h(x,t)$) and (ii) the local polarization (${\bf P}(x,t) = P(\cos \theta(x,t), \sin \theta(x,t))$). In particular, we expect the height field to undergo noisy advection-diffusion dynamics where the advection direction is determined by the ${\bf P}$ field. On the other hand, ${\bf P}$ should follow a dynamics that not only allows spontaneous symmetry breaking ($\partial_t {\bf P} \propto a {\bf P} + b|{\bf P}|^2{\bf P}$) but also contains alignment and anti-alignment terms ($\partial_t {\bf P} \propto D_1\nabla^2{\bf P} + D_2 \nabla^4 {\bf P}$). We expect the transition from the stiff to C-shape to be accompanied by a change in sign of the coefficient of the aligning term $D_1$. How does $h(x,t)$ affect the dynamics of ${\bf P}$? For both phases I and IV of Fig.~\ref{fig_phase}, the local orientation is along the local normal to the interface. In other words, the local orientation is driven by the local curvature ($\nabla^2 h$) of the chain. In effect, we expect such curvature-driven polarization to then introduce non-local forces in the dynamics of the height field. Further, instead of periodic boundary conditions typically used to model growing interfaces, ``clamped" boundary conditions are potentially required for the curvature to set in. These ideas provide challenges for further work. \\

% \subsection{Related work}
% \textcolor{black}{Mention the exponents $z$ and $\alpha$ for each case below including ours. Let us keep this paragraph self-contained. }
It is of further relevance to note where these results - the existence of growth and roughness exponents that are unique to this particular topology - fit in the broader study of active interfaces/membranes. Firstly, it is noted that non-equilibrium extensions to passive interface models have been theoretically investigated with propositions of new universality classes. For instance, by studying capillary fluctuations using a non-equilibrium field theory, it has been shown that (in 2D) $\alpha = 1/2$ and $z=3$, whilst additional non-linear terms decrease these values slightly \cite{fausti2021capillary, besse2023interface, cates2025active}. These interfaces arise from underlying scalar fields derived from conservation laws whose dynamics violate symmetries found in equilibrium. A parallel set of active membrane models have been studied where a fluctuating membrane is coupled to a membrane-spanning protein pump (the concentration field)  that exerts a mechanical force on the membrane upon exchanging material with it (chemical coupling), leading to traveling waves \cite{ramaswamy2000nonequilibrium, ramaswamy2001physics, manneville2001active}. A recent set of works study a variant of this in which the protein dynamics is modeled as inclusions in the height field that move along the membrane with an additional time scale \cite{cagnetta2018active, cagnetta2019statistical, cagnetta2020kinetic}. For these sets of ``active membrane" models, the height correlations are characterized by $S(q) = \langle h(\mathbf{q}) h(\mathbf{-q}) \rangle \sim 1/q^{2}$ universally, the result for a Monge (passive) interface with Gaussian fluctuations \cite{helfrich1973elastic}. This gives rise to the standard $\alpha = 1/2$ roughening. 
For orientational correlations, though these have not been studied elsewhere to our knowledge, perhaps the closest set of works pertain to those studying tangential angular correlations on a liquid--liquid interface undergoing phase separation, where the tangential angular correlations show a smoothening at large scales  \cite{adkins2022dynamics, zhao2024asymmetric}. It is evident that the physics of these aforementioned systems are starkly different from that of the autophoretic C-shape chain; we nevertheless mention them here for the sake of completion. \\

% \subsection{Potential experimental realization}
It remains an open challenge to verify our predicted exponents in an experimental system. To observe significant positional fluctuations in experimental realizations, if one equates damping forces to thermal forces, $\eta b v_s  \sim k_b T/b$ ($\eta$ the viscosity of the medium and $v_s$ the self-propulsion $ \approx 10 \mu m s^{-1}$) (e.g. for droplet experiments \cite{kumar2023emergent, kumar2025temperature, thutupalli2018flow}), one obtains $b  \approx 10^{-7} m$, thus around 10 times smaller than typical chemically interacting colloidal systems.
On the other hand, other phoretic colloidal systems (typically electrophoretic and bubble propulsion mechanisms) have reported propulsion speeds of $v_s \sim 1 \mu m s^{-1}$ \cite{howse2007, ebbens2010pursuit}, where the thermal fluctuations will be appreciable. Constructing extended propelling objects with a well-defined propelling interface out of these units remain an open challenge.
The smaller colloids of $b \sim 10^{-7} m$ deserve a specific consideration. These correspond to so-called ``nanomotor" systems, sub-micron (mostly metallic) colloids whose synthesis routes are well established \cite{zhang2017janus, novotny2020nanorobots} and have been shown to self-propel via various mechanisms \cite{lee2014self, sanchez2015chemically, zhou2021magnetically, song2025polymersome}. Noting that $\tau_r \propto b^{3}$, we find that with $b \sim 10^{-7} m$, $\tau_r \sim 10^{-2} s$, which we note is smaller than the regime required for C-shape formation to take place (c.f. Section \ref{sec:VC}). Thus, it may be feasible to only observe the orientational growth and smoothening with an appreciable $b \sim 10^{-6} m$, where $\tau_r \sim 10^{1} s$, which supports the C-shape.  At this scale it is also important to note that motion is no longer restricted to be in two dimensions, and continuum theories of phoretic/fluid interactions may in addition break down \cite{santiago2018nanoscale, ju2025technology}. In this context it is of interest to note recent works that have reported how phase change in 8CB liquid crystal emulsions (micron-sized) can be induced via external temperature changes \cite{kumar2025temperature}. In the nematic/isotropic phases, the individual colloids no longer merely deterministically self-propel but in addition have a fluctuating propulsion direction. This would correspond to finite $\tau_r$ in our model without sacrificing monomeric size. Creating extended objects out of such systems, and/or realizing a robust C-shape with positional fluctuations thus remain an open problem. Finally, though extended colloidal chains propelled via external (as opposed to autophoretic) mechanisms have widely been synthesized \cite{jayaraman2012autonomous, aubret2018targeted, prathyusha2022emergent, zhang2016natural, nishiguchi2018flagellar, snezhko2011magnetic}, none of them have reported well-defined shapes where interfacial height deviations and fluctuations may be readily computed - i.e. none of these reported shapes propel perpendicular to the body axis \cite{kumar2014actomyosin, subramaniam2024rigid}. It is noted that these systems suffer from the same translational noise suppression and noise scaling issues mentioned above.\\

 % Potential open questions thus naturally arise as to what appropriate (tractable) coarse-grained (continuum) models could describe the active interface presented here. It remains to be seen whether or not both the C-shape and locally flat topologies could be explained by the same continuum model. In addition, instead of periodic boundary conditions typically used to model growing interfaces, ``clamped" boundary conditions are required for the curvature to set in. These ideas provide challenges for further work. \\ 

%In addition, these results could also be relevant to active interfacial systems such as colonies of migrating bacteria \cite{hayakawa2020polar}, where chemical interactions are typically long-ranged. Another open problem, would be the extension of this model to a ring topology, where the possibility of attaining spin waves is imminent \cite{adkins2022dynamics}. 

\section{Appendix}
\appendix
\setcounter{equation}{0} % Reset equation counter
\renewcommand{\theequation}{A\arabic{equation}} % Redefine the equation format

\subsection{Incorporation of chemical trails}
\label{sec:hist_dep}
\begin{figure*}[t!]
    \centering
    \includegraphics[width=0.9\textwidth]{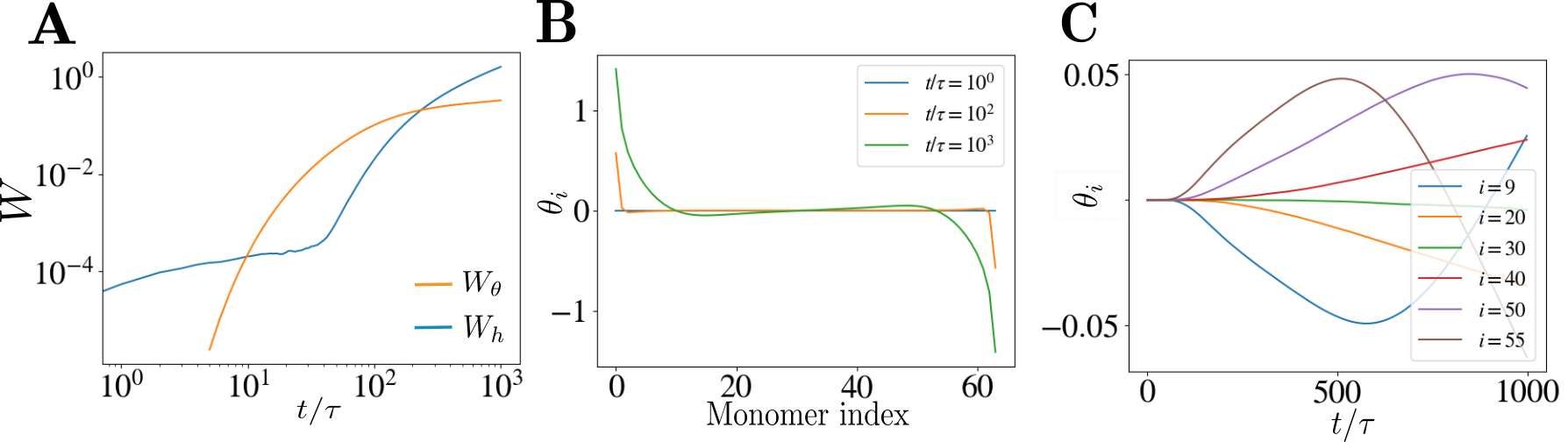}
    \caption{Fluctuations of model with chemical trails. Example of \textbf{A} $W_h$ and $W_{\theta}$ evolution, with the dynamic $\beta_{h}$ labelled (fit between $t/(\tau N^{z_h}) \in [100,200]$) and $t/\tau \in [1,10]$ respectively. \textbf{B} Orientation profile of the C-shape, taken at different time points. \textbf{C} Dynamical evolution of $\theta_i$ for selected monomers $i$.. Here, $D_c=1$, $t_0=1$, with $N=64$ and $T=10^{5}$.}
\label{fig10History}
\end{figure*}

The various results quoted above have been attained via the initial conditions specified in (\ref{eq:init_cond}) (a straight chain with parallel orientation). In this section, we further show that the height fluctuations are \textit{universal} if one adds history-dependence (chemical trails) into the dynamics; indeed these exist in experimentally realizable chemically interacting colloidal systems \cite{jin2017chemotaxis, subramaniam2024rigid}. Thus, instead of Eq.\eqref{eq:chem_field}, for this section, we use the following form of $\mathbf{J}$:
\begin{align}
    \mathbf J_i(\mathbf r_i,t) &= \frac {c_0}{\pi b}
    \sum_{\substack{j=1
    % \\i = j
    }}^N 
    \left[ 
    \int\limits^{t-t_0}_{0} dt'
    \left(
    \frac{\Delta\mathbf r_{ij}  }{\mathcal D^2}
    \right)
    % {\mathcal{F}_{ij}(t,t')}
    \exp \left(
    -\frac{\left[\Delta\mathbf r_{ij}\right]^2  }{\mathcal D}
    \right)
    \right],\nonumber \\
    \Delta\mathbf r_{ij}  &= \mathbf {r}_i(t) - \mathbf r_j(t'),\qquad 
    \mathcal{D} = D_c |t - t' |.
    \label{eq:curr}
\end{align}
 where $t_0$ sets the upper bound on the memory kernel. This corresponds to monomers sensing chemical trails of their neighbors. We display the results in Fig. \ref{fig10History}, for $N=250$ (note in this section $N'=N$). Taking into account the timescale $\tau_c = \frac{b^{2}}{D_c}$ for diffusion of filled micelles across the system, we note that it is necessary to have $\tau_c << \tau_f$ with $\tau_c \sim \tau$ to obtain the C-shape configuration~\cite{kumar2023emergent, subramaniam2024rigid}. 

We present the results for $W_h$ and $W_{\theta}$ in Fig. \ref{fig10History} (A). We observe that the super-ballistic growth regime of $W_h$ is reproduced. With the micellar trails, when one starts such a system with arbitrary initial conditions (e.g. random), the trails deposited via (\ref{eq:curr}) give rise to a forward-backward symmetry breaking in the orientational response to the chemicals. Thus, the system breaks the initial symmetry imposed on it and further universally picks up the C-shape. In our system, we choose sufficiently random initial conditions such that this effect is seen; we then compute $\Delta h$ once the chain has a well-defined interface.\\

For this section only one system size, $N=64$, was studied, averaged over $10$ realizations. Note that the dynamical evolution via (\ref{eq:curr}) requires at each time step a sum over all previous time steps. Thus, even for this system size (on a desktop computer with processor frequency of 2.40GHz and 16 cores), one realization takes $3.77 \times 10^5$ seconds. Thus, a rigorous system-size determination of roughening/smoothening is prohibitive. With only one curve for $W$, obtaining $\alpha$ and $\beta$ via (\ref{eq:fv_scaling}) (\ref{eq:theta_scaling}) requires self-consistency. For $W_h$, we fix $\alpha \approx 0.9$; in this setting the best fit obtained was $\beta_h = 1.41 \pm 0.01$. We have further checked that super-ballistic $\beta_h$ is always obtained if the coarsening is in between evaporative and height conserving limits (this is solely a consequence of the C-shape topology and its fixed bending rigidity, irrespective of dynamical details; see arguments in Section \ref{sec:summ}).\\

Further, for $W_{\theta}$, we find that fixing $\beta_{\theta} = 1$ gives $\alpha_{\theta} \approx 0.2$ -- thus, not a smoothening exponent, but nevertheless much less rough than the standard growth models. If one attempts to fit a negative $\alpha_{\theta}$ instead, one instead finds super-ballistic orientational growth (e.g. $\alpha_{\theta} \approx -0.1$ gives $\beta_{\theta} \approx 4$). Super-ballisticity could also be anticipated given that the time evolution contains higher order temporal derivatives, specifically for monomers near the edges (e.g. $i=9, 50, 55$ in Fig.\ref{fig10History}(C) versus $i=311$ of Fig.\ref{fig4}(D)), thus any fluctuations about the mean will deviate from a simple ballistic scaling. On the other hand, the steady-state angular profile is no longer fully polarized along the propulsion direction (compare Fig.\ref{fig10History}(B) versus Fig.\ref{fig4}(C)), thus $\alpha_{\theta}$ is expected to be higher than in the case without trails.\\

A more complete picture obviously will arise with analyses at various system sizes. However, it is safe to conclude that (i) super-ballistic positional growth; (ii) roughnening of positional fluctuations; and (iii) a roughening of orientational fluctuations that is significantly less that standard models (e.g. EW/KPZ and/or Gaussian membranes) is present when chemical trails are included. \\

% Our conclusions of this section are, thus, that the addition of trail-mediated interactions only appreciably affects the growth of $W_{\theta}$, with the growth and roughness of $W_h$, along with smoothness of $W_{\theta}$ unchanged.\\

\subsection{Early time approximation and probability distribution}
\label{sec:prob_dist}

We write down here the approximate early time dynamics for our model, showing that it represents a driven one dimensional interface and how the diffusive and sub-diffusive scalings for $W^{2}_h(t)$ can be accounted for. Let us explore the regime where orientation dynamics is yet to affect the translational dynamics. Let us re-write the positional dynamics of (\ref{eq:dyn}) for the case of the flat interface, ignoring orientation dynamics. In this case, we have
\begin{align}
    \frac{\partial h_i}{\partial t} = v_s + \mu \Big( |h_{i+1} - h_{i} - 2b | - | h_{i} - h_{i-1} - 2b | \Big) + \xi_{i,t}
\label{eq:app_discr_theta0}
\end{align}
Taking the continuum limit of the above, we have
\begin{align}
    \frac{\partial h}{\partial t} = v_s + \tilde{\mu} \frac{\partial^{2} h}{\partial y^{2}} + \xi
\label{eq:app_cont_ew}
\end{align}
which is thus a standard diffusive interface driven by a propulsion velocity $v_s$ and a Gaussian white noise $\xi$. At very early times, interface fluctuations are dominated by $\xi$, before the diffusive term comes into play.\\

These height fluctuations are alternatively characterized by the distribution of fluctuations, $P(W^{2})$ \cite{takeuchi2012evidence, Antal95}. For the standard EW interface (non-propelling), known results exist and are re-written here. We can directly quote the result from \cite{Antal95} for arbitrary initial height distribution

\begin{equation}
  P(W^2, t) = \int_{-i \infty}^{i\infty} \frac{dy}{2\pi i} e^{xy} \prod_{n=1}^{\infty} \frac{-y s^2_{n0}e^{-n^2 \tau}/(1+ y a_n)}{  \langle W^2 \rangle_{SS} 
  \left(1+y a_n\right)},\nonumber
\end{equation}
where $W^2_{SS}=\frac{N D_t}{12 \nu_h}, a_n = \frac{6}{(\pi n)^2}(1-e^{-\tau_{p} n^2})$, and the scaling variables are given by
\begin{equation}
    x= \frac{W^2}{ W^2_{SS}}, ~ \tau_{p} = \frac{8 \pi^2 \nu_h t}{L^2}, ~ s_{n0}^2 = \frac{2|a_n(0)|^2}{ W^2_{SS}} \nonumber.
\end{equation}
The terms $a_n(0)$ are Fourier transforms of the initial height profile. In our case, we start with $h(x, t=0)=0$, which fixes $a_n(0)=0$. For a flat initial height profile, one has $s_{n0}=0$, which gives \cite{Antal95}:
\begin{equation}
    P(W^2, t) = \sum_{m=1}^{\infty} \frac{1}{a_m} \frac{\exp\left[ \frac{-x}{a_m}\right]}{ \langle W^2 \rangle_{SS}} \prod_{n=1, \neq m}^{\infty} \frac{a_m}{a_m- a_n}  .
\label{eq:app_dist_exact}
\end{equation}

\begin{figure}[b]
    \centering
    \includegraphics[width=0.44\textwidth]{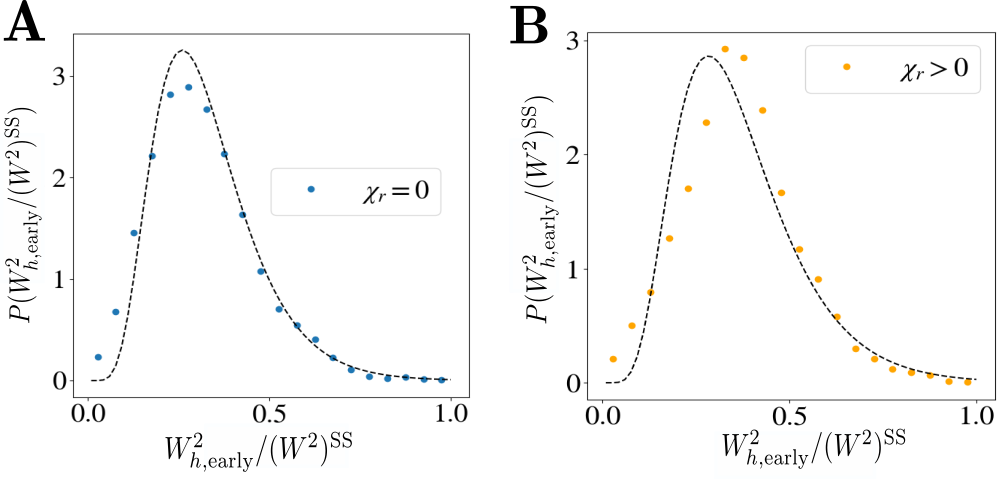}
    \caption{\textbf{A}: Distribution of height fluctuations at early times ($t\ll \tau_f$). The distribution of $W^{2}_h$ of the non-interacting chain at early times (blue, scatter) can be fitted to the known exact solution for the height distribution of the EW interface; here the fitted $\tau_p  \approx 0.17$. \textbf{B}: The chemically interacting chain (orange scatter) shows a deviation from this, with the best fit of $\tau_p  \approx 0.21$ again in dashed. The average of least squared errors for \textbf{A} and \textbf{B} are $0.0508$ and $0.0728$ respectively, with the latter indicating an enhanced deviation from the EW distribution. Parameters chosen are as follow: for \textbf{A}, $\Lambda \rightarrow \infty$, whilst for \textbf{B}, $\Lambda = 0.2$. In both cases $\tilde{\Lambda} = 100$. }
\label{fig06PDF}
\end{figure}

The early time approximation of (\ref{eq:app_cont_ew}) corresponds to setting $\Lambda \rightarrow \infty$. Note that one can either probe $\tau << \tau_f$ or $\chi_r = 0$ globally; we pick the latter. The analytical distribution of (\ref{eq:app_dist_exact}) can then be compared to this limit; we find that the distribution is reproduced - black line and blue scatter of Fig. \ref{fig06PDF}(A). We can thus suitably compare this with the case of $\Lambda$ finite, restricting to $\tau << \tau_f$. Here, we find that our propelled interface displays deviations from this passive distribution; this is displayed in Fig. \ref{fig06PDF}(B) - yellow scatter. For both cases, we plot alongside the best fit analytical distribution in dashed black, and we further find that the averaged least squared error of the fit is substantially larger in the case of $\Lambda$ finite; we thus conclude that the early time non-equilibrium distribution deviates from the passive EW case. This marginal deviation at early times arises again via the aforementioned roto-translational coupling; this thus again constitutes a non-equilibrium signature that appears via deterministic (but small) orientational interactions. This deviation in the height distribution is analogous to that seen in so-called ``anomalous diffusion" \cite{wang2009anomalous, wang2012brownian}, where microscopic colloidal particles in specific systems display a mean-squared displacement exponent of $1$, but the probability distribution of displacements has been found to follow the Laplace (instead of Gaussian) distribution \cite{wang2012brownian, chechkin2017brownian}. The exact analytical distribution of the chemically interacting chain thus remains an open future problem.

\subsection{Linearised approximation of super-ballistic growth exponent}
\label{sec:app_super_ball}
In the main text we have reported the growth exponent of the C-shape to be $\beta_h \sim 1.7$ (super-ballistic). Although rationalizing this for a curved surface is beyond exact analytical calculation in this work; we study instead a flat interface, via a solution to the approximate linearised dynamics.\\

Let us study the simplest case of roto-translational coupling, in particular, where (small) angular dynamics set in; in this setting the full set of equations are solvable.
 Consider the $i$th monomers' positional dynamics from (\ref{eq:dyn}). For small $\theta_i$, we can write this as
\begin{align}
    \dot{x}_{i} = v_s \left(1 - \frac{\theta_i^{2}}{2}\right) + \mu F^{b}_{x_i} + \xi_{t,i}
\label{eq:app_ydyn}
\end{align}
Note that the $x_i$ are assumed to be stationary in this regime (flat interface). We can use this for the angular dynamics to write
\begin{align}
    \dot{\theta}_i = \chi_r \left(1 - \frac{\theta_i^{2}}{2}\right) S_{i}  + \xi_{r,i}
\label{eq:app_thetadyn}
\end{align}
where $S_i = \frac{1}{4b^{3}}\sum_j{\frac{1}{(i-j)^{2}}\Big[2\Theta(i-j)-1 \Big]}$ for a completely flat chain.  Here, $\Theta(x)$ is the step function. This simplification allows us to write a linearized (\ref{eq:app_thetadyn}), without accounting for chemical currents along the $x$ direction. We note that fluctuations along the body axis ($y$ direction) mean that $S_i$ is random quantity at each time step. The deterministic solution to this is
\begin{align}
    \theta_i = \sqrt{2} \frac{1 - \exp[-\chi_r S_i t]}{\exp[-\chi_r S_i t] + 1}
\label{eq:app_thetasol}
\end{align}
We next probe the early time regime of $t << \tau_f$; this corresponds to the approximation of a flat surface (c.f. Fig. \ref{fig_phase}(d)). We then plug (\ref{eq:app_thetasol}) into (\ref{eq:app_ydyn}). We then use the small argument expansion for $\cosh(z) \approx 1 + \frac{z^{2}}{2}$, with $z = \frac{\chi_r S_i t}{2}$. This gives us 
\begin{align}
    \dot{x}_i = F_i(t) \approx v_s \Big(1 -  \frac{b^6}{2\tau_f^2} S_i^{2} t^2 \Big) + F^{b}_{x,i} + \xi_{t,i}
\label{eq:early_ydot}
\end{align}
thus
\begin{align}
    \langle X \rangle = v_st - \frac{b^6}{6\tau_f^{2}} \langle S^{2}\rangle t^3
\label{eq:app_soly}
\end{align}
where $\langle \rangle$ is performed over both the monomers and realizations. For a flat chain (no displacements along propulsion direction), $S_i$ is given by the deterministic formula above. In general, Here, $X$ is a site-independent average of the chain position. We obtain the scaling for $(\Delta h)^{2}$ as follows
% \begin{subequations}
\begin{align}
    % \langle W_h^{2} \rangle &\approx \int \langle F^{b}(t') F^{b}(t'') \rangle dt' dt'' + \int \langle \xi(t') \xi(t'') \rangle dt' dt'' 
    % + \frac{v_s^{2} \chi_r \langle S \rangle^{2}}{4} t^{4} - \frac{v_s \chi_r \langle S \rangle}{2}\Big( \int \langle F(t') \rangle dt'\Big) \nonumber\\
    % &= v_s^{2} \chi_r^{2} \frac{\langle S^{2} \rangle}{3} t^3 + 2D_tt
    W_h^{2} &\approx \frac{b^{6}}{6\tau_f^{2}} \langle \Big(S_i^{2} - \langle S \rangle^{2} \Big)^{2} \rangle t^6 + 2D_tt
\label{eq:h2_approx}
\end{align}
% \end{subequations}
where we have used that $\langle F^{b}(t') \rangle = 0$ and $\langle F^{b}(t') F^{b}(t') \rangle = \langle F^{b}(t) \rangle \langle F^{b}(t') \rangle$. Strictly speaking, the latter will cease to be true when curvature sets in (spring forces then become correlated), but it will suffice for the flat approximation used here. We thus obtain the result that the dynamics is super-ballistic with exponent $\sim 3$ (with an early time diffusive $\sim 0.5$). Here, fluctuations along the body axis are minimally accounted for such that $\text{Var}(S) \neq 0$. We thus conclude that minimal roto-translational coupling on a flat interface can predict diffusive to super-ballistic growth transitions but clearly overestimates the super-ballistic exponent to that of a surface with finite curvature.

\subsection{Early-time approximation to ballistic orientational growth}
The deterministic solution on a quasi-flat chain also enables us to rationalize the $\beta_{\theta} \sim 1$ at early times. From (\ref{eq:app_thetasol}), let us probe $t<<\tau_f$, where we find
\begin{align}
    \theta_i \approx \frac{1}{\sqrt{2}}(\chi_r S_i t)[1 + \frac{\chi_r S_i t}{2}]
\label{eq:app_theta_early}
\end{align}
Next, we exploit the symmetries of the orientations along the body axis (c.f. Fig. \ref{fig4}(c)), where we find that $\langle \theta \rangle_i = 0$ and by extension $\langle (\Delta \theta)^{2} \rangle = \langle (\theta_i - \langle \theta \rangle_i)^2\rangle = \langle \theta_i^2 \rangle$. Thus
\begin{align}
    \langle (\Delta \theta)^2 \rangle &= \langle \theta_i^{2} \rangle = \frac{b^6}{2\tau_f^2} \langle S_i^{2} \rangle t^{2} + O(t^{3}) + O(t^{4}) 
\label{eq:app_msd_orient_approx}
\end{align}
where the terms $O(t^{3,4})$ are sub-leading for $t << \tau_f$. Thus, a combination of the symmetry of the orientations along the chain ($\theta_i = -\theta_{N-i}$) and the transverse fluctuations approximation of the previous section gives us $\beta_{\theta} \sim 1$ at early times.    

\subsection{Curvature of chain}
\label{sec:app_curvature}
To calculate the phase diagram in Fig. \ref{fig_phase}, we use the Monge representation to calculate the curvature in the steady-state \cite{helfrich1973elastic}, following \cite{subramaniam2024rigid}. We compute the curvature as:
\begin{align}\label{curv_def}
  \kappa = \left\langle  \nabla \cdot \left[ {\nabla h_{i}} / {\sqrt{1 + | \nabla h_{i} |^{2}}} \right]  \right\rangle,
% $%
\end{align}
where $h_{i}$ is the height function of the chain, evaluated at each monomer location, measured from the vertical line connecting the edge monomers. The average is performed across the monomers in the chain, giving a scalar value. In our case, the inner bracket is simply reduced to $\partial_{y} \left[ {\partial_{y} x_i(y)}/{\sqrt{1 + | \partial_{y} x_i |^{2}}} \right]$.
For Fig. \ref{fig_phase}, we take the averaged $|\kappa|$ in the steady state. 
%Further, for selected values of $N$, we plot the C-shape curvature in Fig. \ref{fig_app_curv}. 
Further, we note that $\kappa$ $\rightarrow 0$ as $N \rightarrow \infty$ \cite{subramaniam2024rigid}. For a circular interface, $\kappa$ would be scale as $1/R$ ($R$ the radius of the circle), whereas for a flat interface $\kappa$ would be constant. Thus, this variation of the $\kappa$ profile can be used as proxy definition for the C-shape. 

%%%================
%%%================
\begin{figure} 
    \centering
    \includegraphics[width=0.457\textwidth]{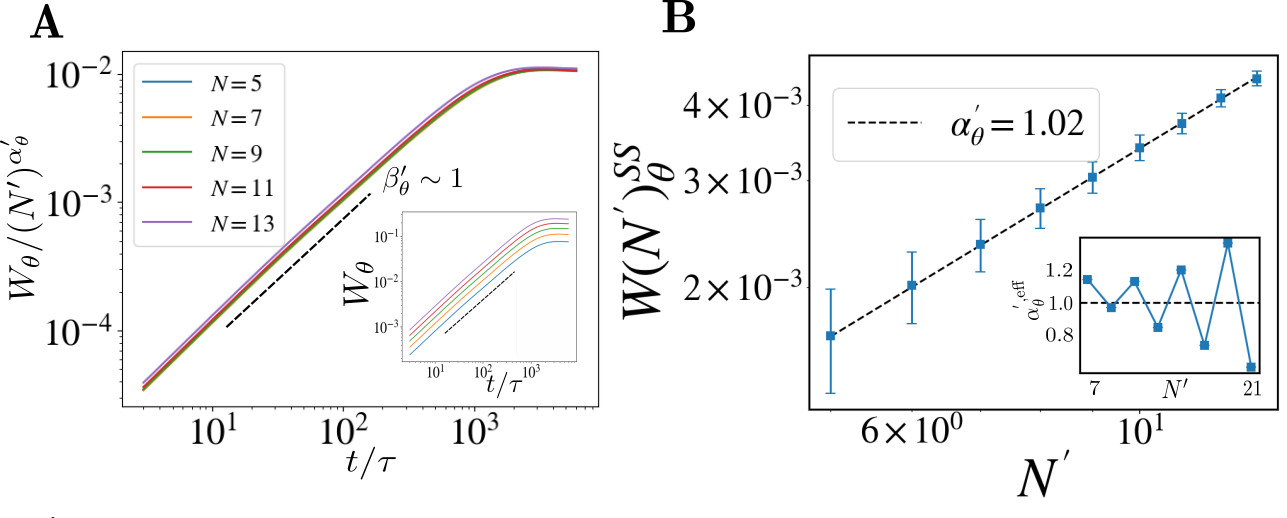}
        %\caption{Lorem ipsum}
    \hfill
    \caption{Scaling of orientational fluctuations in locally flat region. \textbf{A} $W_{\theta}$ for different locally flat lengths $N'$, with the dynamic $\beta_{\theta}'$ labelled (fitting window identical to Fig. \ref{fig4}). The same scaling is plotted in the \textit{inset}. \textbf{B} Roughness scaling for $W_{\theta}^{\text{SS}}(N')$. Here, $N=312$ and $\tau_f = 0.2$, $\Lambda=0.2$, $\tilde{\Lambda}=0.01$.}
\label{fig8ThetaF}
\end{figure}
%%%================
%%%================
\begin{figure} [b]
    \centering
    \includegraphics[width=0.4\textwidth]{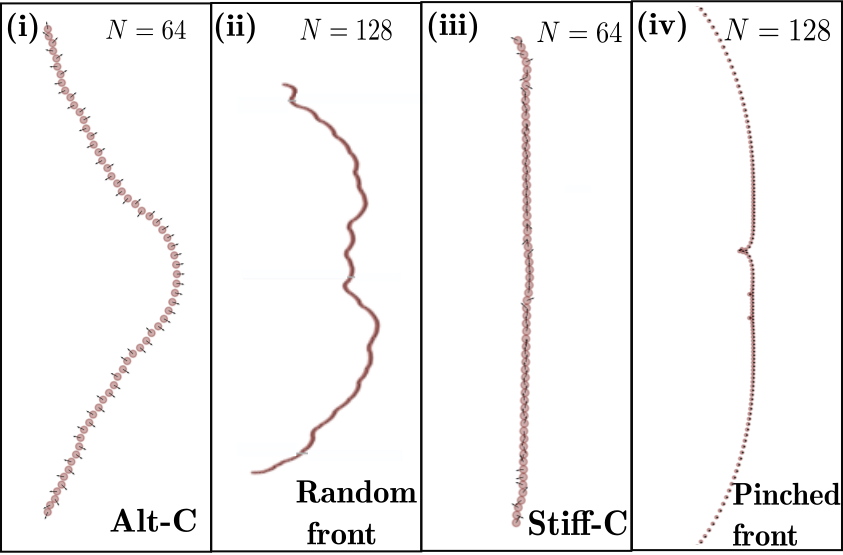}
        %\caption{Lorem ipsum}
    \hfill
    \caption{Examples of DSS. From left to right: (i) Alt-C, (ii) Random front, (iii) Stiff-C, and (iv) Pinched front respectively. $\{ \Lambda, \tilde{\Lambda} \}$ values are (i) $\{ 10^{-5},2\}$, (ii) $\{ 0.05,0.2\}$, (iii) $\{ 10^{-6},200\}$, (iv) $\{ 10^{1},200\}$. Note that (iv) in addition incorporated trails via (\ref{eq:curr}); here $t_0=1$ and $\tau_r = 5$ in addition. }
\label{fig_app_dss}
\end{figure}
%%%================
%%%================

\subsection{Scaling of $W_{\theta}$ in the locally flat regime}
\label{sec:app_wtheta_flat}
The scalings of $W_{\theta}$ in for $N'<<N$ is presented in Fig. \ref{fig8ThetaF}. We find that the scaling law of Eq. (\ref{eq:theta_scaling}) is satisfied; the ballistic $\beta_{\theta}'=1$ is retained whilst the roughness 
$\alpha_{\theta}'\approx1.02\pm02 $ is obtained. Note that this is opposed to the smoothness exponent obtained on the full C-shape. We also note that the specific choice of $N'\in [ 5,21 ]$ used here, thus at most $\approx 4 \%$ of the chain length, is substantially smaller than in Fig. \ref{fig5}. For any appreciably larger $N'$ the scaling of (\ref{eq:theta_scaling}) is not observed. These results thus further highlight the fact that the smoothness exponent found in Fig. \ref{fig4} is a sole consequence of the C-shape; with distinct topology-specific results vis-a-viz a flat surface. 
\subsection{Other DSS for large $N$}
\label{sec:app_dss}
It is to be noted also that the phase diagram in Fig. \ref{fig_phase} being $N$ specific and only comparing ($\Lambda$, $\tilde{\Lambda}$) does exclude other DSSs that arise from simulations. We show four examples in Fig. \ref{fig_app_dss}, where we label them as (i) ``Alt-C", (ii) ``Random front", (iii) ``Stiff-C", and (iv) ``Pinched front", respectively. We note that DSS (ii) corresponds to the ``Disordered" phase in Fig. \ref{fig_phase}.  For each of these DSSs, the chain propels in a deterministic direction along the x-axis; thus (\ref{eq:d2h}) can be computed in principle. However, none of these DSSs satisfies all the four criteria enumerated in Sec. \ref{sec:condition_Cshape} to obtain a stable C-shape. We see that for (i) criteria 2, 3, and 4, for (ii) 1 and 3, for (iii) 1,3 and 4; and for (iv) 2 and 3 are satisfied. Thus, the C-shape is the unique topology that satisfies all the four criteria. The fluctuations for $W_{\theta}$ for each of these shapes can also, in principle, be computed, but we do not pursue such questions here. \\

% \subsection{Fixed segment scaling}
% For a fixed segment length $N'$, we can further show that the roughness exponent is equal to the locally flat roughness exponent. As explained in Section ?? previously, this equivalence is attributed to the fact that at large $N$, the chain asymptotes to a straight line (see Fig ???).

% \begin{figure*}[ht!]
%     \centering
%     \includegraphics[width=0.65\textwidth]{figures/FIG_APP_N'FIX.png}
%         %\caption{Lorem ipsum}
%     \hfill
%     \caption{FV scaling versus $N$ for a fixed $N' = \frac{N}{4}$. \textbf{A}: .}
% \label{fig_app3}
% \end{figure*}

% \subsection{Effect of deviation from }

%\subsection{Effect of $\tau_r$}

\section*{Acknowledgments}
We thank Professors ME Cates and M Muthukumar for useful discussions. 
We also thank the anonymous referee for their feedback and suggestions, which led to
an improvement in the presentation of our results. 
AGS acknowledges funding from the DIA Fellowship from the Government of India.
TB is supported through the Luxembourg National Research Fund (FNR), grant reference C22/MS/17186249.

%%----------
%\bibliography{references}

%

%%%---------------------------
%%%%================
%%%%================
\end{document}